\documentclass[12pt]{article}
\usepackage{amssymb}
\usepackage{amsfonts}
\usepackage{graphicx}
\usepackage{dcolumn}
\usepackage{bm}
\usepackage{amsmath}
\usepackage{amsfonts}
\usepackage{amssymb}
\usepackage{amsmath}
\usepackage{amsfonts}
\usepackage{color}
\usepackage{hyperref}
\begin{document}

\begin{center}
{\Large\bf Continuum Random Phase Approximation for Relativistic Point Coupling Models}
\end{center}
\vspace{0.5cm}
\begin{center}
\long\def\symbolfootnote[#1]#2{\begingroup%
\def\thefootnote{\fnsymbol{footnote}}\footnote[#1]{#2}\endgroup}

{\large J. Daoutidis\symbolfootnote[1]{Electronic address: jdaoutid@ph,tum.de}, P. Ring\symbolfootnote[2]{Electronic address: ring@ph,tum.de}}
\vspace{0.5cm}

{\em Physik-Department der Technischen Universit\"at M\"unchen, D-85748 Garching, Germany}.\\
\end{center}
\date{\today}
\hspace{1.0cm}

\begin{abstract}
Relativistic Continuum Random Phase Approximation (CRPA) is used to investigate collective excitation phenomena in several spherical nuclei along the periodic table. We start from relativistic mean field calculations based on a covariant density functional with density dependent zero range forces. From the same functional an effective interaction is obtained as the second derivative with respect to the density. This interaction is used in relativistic continuum-RPA calculations for the investigation of isoscalar monopole, isovector dipole and isoscalar quadrupole resonances of spherical nuclei. In particular we study the low-lying E1 strength in the vicinity of the neutron evaporation threshold. The properties of the resonances, such as centroid energies and strengths distributions are compared with results of discrete RPA calculations for the same model as well as with experimental data. \end{abstract}
PACS numbers: 21.30.Fe, 21.60.Jz, 21.65.+f, 21.10.-k

\bigskip

\section{Introduction}

Density functional theory (DFT) provides a very successful
description of nuclei all over the periodic table.
Based on relatively simple functionals, which are adjusted in a
phenomenological way to the properties of infinite nuclear matter and
a few finite nuclei, this theory allows a highly accurate
reproduction of many nuclear structure data, such as binding
energies, radii, deformation parameters of finite nuclei and their
dependence on mass number and isospin. In addition to these static
properties, one can use the nuclear response to external multipole
fields to investigate the dynamics of such systems. In the framework
of time dependent density functional theory, this response can be
calculated from the linearized Bethe Salpeter equation using an
effective interaction derived from the same functional.

A very successful scheme of this type is covariant density functional theory
(CDFT). It is based on Lorentz invariance, connecting in a
consistent way the spin and spatial degrees of freedom of the nucleus.
Therefore, it needs only a relatively small number of parameters which are
adjusted to reproduce a set of bulk properties of spherical closed-shell
nuclei. Numerous works have shown that observations involving both ground
state and excited state phenomena, can be nicely interpreted in a relativistic framework.

The most popular applications of this type are based on the Walecka
model~\cite{SW.86}, where the nucleus is described as a system of
Dirac nucleons interacting with each other via the exchange of
virtual mesons with finite mass and the electromagnetic field through
an effective Lagrangian. In the mean field approximation this yields
to various contributions to the nuclear self energy depending on the
quantum numbers of these mesons. Early investigations have shown that this simple ansatz is not able to describe the incompressibility of infinite nuclear matter nor the surface properties of finite nuclei such as nuclear deformations. For that reason, a medium dependence has been introduced by including nonlinear meson self-interaction terms in the Lagrangian~\cite{BB.77}.

Several very successful phenomenological RMF interactions of this type have been adopted, as for instance the popular set NL3~\cite{NL3}. Closer to the concept of density functional theory are models with an explicit density dependence for the meson nucleon couplings. This density dependence can be calculated from first principles in a microscopic  Dirac-Brueckner scheme~\cite{FLW.95} or it can be adjusted in a completely phenomenological way to properties of finite  nuclei~\cite{TW.99,DD-ME2}.

One of the advantages of density functional theory is the fact that
with a proper choice of the parameters the success of RMF for nuclear
ground states ensures also a good basis upon which one can apply
time-dependent density functional theory to study nuclear
excitations. In order to investigate the dynamic behavior of the
nuclear system, one considers oscillations around the self-consistent
static solution. This can be done by solving the time dependent
relativistic mean field equations (TDRMF)~\cite{VBR.95} or, in the
limit of small amplitudes, by using the relativistic random phase
approximation (RRPA)~\cite{RMG.01}. The corresponding eigen modes can
be determined either by diagonalizing the RRPA equation in an
appropriate basis or by solving the linear response equations in a
time-dependent external field. This requires a matrix inversion for
given frequency $\omega$.

These two methods lead in principle to exactly identical results.
There are, however, cases where one of them is clearly preferable.
The proper treatment of the coupling to the continuum is such a case,
which can be solved in a very elegant way, by the solution of the
Bethe Salpeter equation within the response formalism.

We recall that the spectrum of the Dirac equations has a discrete and
a continuous part. For the ground state properties of the nucleus,
one needs only the single particle wave functions of the occupied
orbitals in the Fermi sea. They are either determined by solving the
corresponding differential equations in $r$-space or by expansion in
an appropriate basis, given for for instance by a finite number of
eigenfunctions of a harmonic oscillator~\cite{GRT.90} or of a
Saxon-Woods potential in a finite box~\cite{ZMR.03}. For the bound
states both methods yield the same solutions with high accuracy.
However, this is no longer true for the states in the continuum. Here
we have, in the first case scattering solutions in $r$-space for each
energy with proper boundary conditions, while in the second case, a
finite number of discrete eigenstates which depend strongly on the
dimension of the expansion. They provide only a basis and have little
to do with physics.

These discrete eigenstates lead to a finite number of $ph$-configurations for
the solution of response equations. with a discrete spectrum. They provide us
with the so called \textit{spectral representation} of the response function
in contrast to the \textit{continuum representation}, where the exact
scattering states with the proper boundary conditions are used at each energy.

   Self-consistent relativistic RPA (RRPA) calculations have a long history. The early investigations in the eighties~\cite{Fur.85,HG.89,SRM.89,DF.90} were based on the Walecka model with linear meson-nucleons couplings. They were able to describe the low-lying negative-parity excitations in $^{16}$O by the method of matrix diagonalization~\cite{Fur.85}, isoscalar giant resonances in light and medium nuclei~\cite{HG.89} by the solution of the linear response equation in the spectral representation, and the longitudinal response for quasi-elastic electron  scattering with a proper treatment of the continuum.

The first RRPA calculations based on non-linear
models were carried out in the spectral
representation including only normal particle-hole ($ph$) pairs with
particles above the Fermi energy and holes in the Fermi sea. This
seemed to be a reasonable approximation, since the configurations
formed by particles in the Dirac sea and holes in the Fermi sea
($ah$-pairs) are more than 1.2 GeV away from the normal $ph$-pairs.
Indeed, a proper coupling to the Dirac sea and current conservation
was neglected in these investigations. They showed considerable
deviations from the results obtained form time-dependent
RMF-calculations with the same Lagrangian, particularly for isoscalar
excitations~\cite{VRL.99}. A fully self-consistent treatment
with current conservation requires the inclusion of a very large
number of $ah$-pairs connected with a considerable numerical effort.
Most of the very successful applications of RRPA theory based on
non-linear meson-nucleon coupling models in the last ten years have
been carried out in this way \cite{VWR.00,MGW.01,VPR.02,PNVR.05,NVR.05}.%

There are also relativistic continuum RPA calculations based on the
non-spectral representation of the response function using the single
particle Green's function in the continuum with proper boundary
conditions~\cite{SRM.89}. These calculations are done for
meson exchange forces with finite range. The early investigations
were based on linear models. Later on the method was
generalized to include non-linear coupling terms between the
mesons~\cite{Pie.00}. This leads to a a more sophisticated density
dependence which is crucial for a realistic description of giant
resonances in nuclei~\cite{Pie.00,Pie.01}.

Of course, because of the finite range of the effective force these models are
relatively complicated not only for static applications to triaxially deformed
or rotating nuclei, but also for investigations of nuclear dynamics, such as
the solution of the relativistic RPA or linear response equations for the
description of excited states. In particular one needs simpler forces for
applications going beyond the mean field approach such as Particle Vibrational
Coupling (PVC)~\cite{LRT.08} or configuration mixing
calculations in the framework of the Generator Coordinate Method
(GCM)~\cite{NVR.06}. Therefore over the years several
attempts have been made to develop relativistic point coupling (PC) models
with forces of zero range~\cite{MM.89}, in analogy to
non-relativistic Skyrme-functionals. but only recently parameter sets have
been found, which are comparable in quality to the density dependent
meson-exchange models~\cite{BMM.02,NVLR.08b}.

PC models do not contain mesonic degrees of freedom and are therefore
closer to the philosophy of the density functional theory. Their
essential advantage is of course the fact that the zero range of the
effective interaction reduces considerably the numerical effort in
practical applications. Because of their simplicity they are nowadays
much used in many complex calculations going beyond the mean field
approach~\cite{LRT.08,NVR.06}.
However, so far they have not been used much for the dynamic
investigations and it is only quite recently that a code has been
developed to diagonalize the RPA equations for relativistic Point
Coupling models~\cite{NVR.05} and it has been shown that this latter
approach reproduces excitation and collective phenomena, in
particular Giant Multipole Resonances, with a quality comparable to
that of standard finite-range forces.

This manuscript is devoted to an investigation of relativistic point
coupling models with an exact treatment of the coupling to the
continuum. The relativistic response equations are solved both in the
continuum and in the spectral representation and the corresponding
results are compared. We use the Lagrangian PC-F1~\cite{BMM.02},
which is capable of reproducing a wide range of experimental data.

The paper is organized in the following way: In Sec.~\ref{PCRMF}
we present the main characteristics of the point coupling RMF theory,
while the relativistic RPA equations are derived in Sec.~\ref{CRPA}.
The proper treatment of the continuum in connection with point
coupling models is discussed in Sec.~\ref{continuum} and in
Sec.~\ref{MGR} we finally present applications of this method for the
spectra of in spherical nuclei. In particular we calculate the
strength function of Isoscalar and Isovector Giant Resonances as well
as their contributions to their respective energy weighted sum rules.
The results are summarized in Sec.~\ref{summary}.

\section{Relativistic mean field theory of zero range.}

\label{PCRMF}

As in all the relativistic models, the nucleons are described as
point like Dirac particles. In contrast to the Walecka
model, however, where these particles interact by the
exchange of effective mesons with finite mass, point coupling
models~\cite{MM.89} neglect mesonic degrees of
freedom and consider only interactions with zero range. In principle,
these models are similar to the Nambu Jona-Lasinio model
\cite{NJL.61a} used extensively in hadron physics. There is, however,
an important difference: in order to obtain a satisfactory description
of the nuclear surface properties one needs gradient terms in the
Lagrangian simulating a finite range of the interaction.

A general point-coupling effective Lagrangian is constructed to be consistent
with the underlying symmetries of $QCD$ (e.g., Lorentz covariance, gauge
invariance, and chiral symmetry). It should in principle contain every
possible term, allowed by these symmetries, but at the same time should also
be described by the least possible number of parameters in order to give a quantitative solution.

In this work we use the point coupling Lagrangian introduced by
Buervenich et al. in Ref.~\cite{BMM.02}. It presents an expansion in
powers of the nucleon scalar, vector and isovector-vector densities.
The Lagrangian
\begin{equation}
\mathcal{L}= \mathcal{L}_{\mathrm{free}}+\mathcal{L}_{\mathrm{4f}%
}+\mathcal{L}_{\mathrm{hot}}+\mathcal{L}_{\mathrm{der}}+\mathcal{L}%
_{\mathrm{em}}%
\label{Lag-PC}%
\end{equation}
consists of the term for free nucleons:
\begin{equation}
\mathcal{L}_{\mathrm{free}}=\bar{\psi}(i\gamma_{\mu}\partial^{\mu}-m_{N})\psi,
\label{Lparts}%
\end{equation}
the term for normal four-fermion interactions
\begin{align}
\mathcal{L}_{\mathrm{4f}}=  &  -\frac{\alpha_{S}}{2}(\bar{\psi}\psi)(\bar
{\psi}\psi)-\frac{\alpha_{V}}{2}(\bar{\psi}\gamma_{\mu}\psi)(\bar{\psi}%
\gamma^{\mu}\psi)\label{L_4f}\\
&  -\frac{\alpha_{TS}}{2}(\bar{\psi}\vec{\tau}\psi)(\bar{\psi}\vec{\tau}%
\psi)-\frac{\alpha_{TV}}{2}(\bar{\psi}\vec{\tau}\gamma_{\mu}\psi)(\bar{\psi
}\vec{\tau}\gamma^{\mu}\psi),\nonumber
\end{align}
the term for higher order terms leading in mean field approximation
to a density dependence
\begin{equation}
\mathcal{L}_{\mathrm{hot}}=-\frac{\beta_{S}}{3}(\bar{\psi}\psi)^{3}%
-\frac{\gamma_{S}}{4}(\bar{\psi}\psi)^{4}-\frac{\gamma_{V}}{4}[(\bar{\psi
}\gamma_{\mu}\psi)(\bar{\psi}\gamma^{\mu}\psi)]^{2}, \label{L_hot}%
\end{equation}
the term containing derivative terms which simulate in a simple way
the finite range of the forces:
\begin{align}
\mathcal{L}_{\mathrm{der}}  &  =-\frac{\delta_{S}}{2}(\partial_{\mu}\bar{\psi
}\psi)(\partial^{\mu}\bar{\psi}\psi)-\frac{\delta_{V}}{2}(\partial_{\mu}%
\bar{\psi}\gamma_{\nu}\psi)(\partial^{\mu}\bar{\psi}\gamma^{\nu}%
\psi)\nonumber\\
&  -\frac{\delta_{TS}}{2}(\partial_{\mu}\bar{\psi}\vec{\tau}\psi
)(\partial^{\mu}\bar{\psi}\vec{\tau}\psi)\label{L_der} -\frac{\delta_{TV}}{2}(\partial_{\mu}\bar{\psi}\vec{\tau}\gamma_{\nu}%
\psi)(\partial^{\mu}\bar{\psi}\vec{\tau}\gamma^{\nu}\psi)
\end{align}
and finally the electro-magnetic part%
\begin{equation}
\mathcal{L}_{\mathrm{em}}=-\frac{1}{4}F_{\mu\nu}F^{\mu\nu}-\frac{e}{2}%
(1-\tau_{3})A_{\mu}\bar{\psi}\gamma^{\mu}\psi.
\end{equation}
In these equations, $\psi$ represents the nucleon spinors. The subscripts $S$
and $V$ are attributed to scalar and vector fields, while the subscript $T$ is
attributed to isovector fields. As usual, vectors in isospin space are denoted
by arrows, where symbols in bold indicate vectors in ordinary
three-dimensional coordinate space.

From this Lagrangian and the corresponding energy momentum tensor we
can
derive a relativistic energy density functional. It has the form:%
\begin{equation}
\mathcal{E}_{\mathrm{RMF}}[\hat{\rho},t]=\int d^{3}r~{H(\bm{r},t}),
\label{Energy}%
\end{equation}
where the energy density
\begin{equation}
H(\bm{r},t)=H_{\mathrm{kin}}(\bm{r},t)+H_{\mathrm{int}}%
(\bm{r},t)+H_{\mathrm{em}}(\bm{r},t) \label{energy_density}%
\end{equation}
consists of a kinetic part
\begin{equation}
H_{\mathrm{kin}}(\bm{r},t)=\sum_{i}^{A}\,{\bar{\psi}_{i}(\bm{r},t)\left(
\bm{\alpha}\bm{p}+\beta m-m\right)  \psi_{i}(\bm{r},t)}, \label{kinetic}%
\end{equation}
an interaction part
\begin{align}
H_{\mathrm{int}}(\bm{r},t)  &  =\frac{\alpha_{S}}{2}\rho_{S}^{2}+\frac
{\beta_{S}}{3}\rho_{S}^{3}+\frac{\gamma_{S}}{4}\rho_{S}^{4}+\frac{\delta_{S}%
}{2}\rho_{S}\Delta\rho_{S}\nonumber\label{E12}\\
&  +\frac{\alpha_{V}}{2}j_{\mu}j^{\mu}+\frac{\gamma_{V}}{4}(j_{\mu}j^{\mu
})^{2}+\frac{\delta_{V}}{2}j_{\mu}\triangle j^{\mu}\\
&  +\frac{\alpha_{TV}}{2}\vec{j}_{TV}^{\mu}\cdot\vec{j}_{TV\mu}+\frac
{\delta_{TV}}{2}\vec{j}_{TV}^{\mu}\cdot\triangle(\vec{j}_{TV})_{\mu}\nonumber
\end{align}
and an electromagnetic part
\begin{equation}
H_{\mathrm{em}}(\bm{r},t)=\frac{1}{4}F_{\mu\nu}F^{\mu\nu}-F^{0\mu}\partial
_{0}A_{\mu}+eA_{\mu}j_{p}^{\mu}.
\end{equation}
The interaction part depends on the local densities:
\begin{align}
\rho_{S}(\bm{r},t)  &  =\sum_{i}^{A}\bar{\psi}_{i}(\bm{r},t)\psi
_{i}(\bm{r},t),\label{densities}\\
\rho_{V}(\bm{r},t)  &  =\sum_{i}^{A}\bar{\psi}_{i}(\bm{r},t)\gamma_{0}\psi
_{i}(\bm{r},t),\\
\rho_{TS}(\bm{r},t)  &  =\sum_{i}^{A}\bar{\psi}_{i}(\bm{r},t)\tau_{3}\psi
_{i}(\bm{r},t),\\
\rho_{TV}(\bm{r},t)  &  =\sum_{i}^{A}\bar{\psi}_{i}(\bm{r},t)\tau_{3}
\gamma_{0}\psi_{i}(\bm{r},t)
\end{align}
and currents
\begin{align}
j_{V}^{\mu}(\bm{r},t)  &  =\sum_{i}\bar{\psi}_{i}(\bm{r},t)\gamma^{\mu}%
\psi_{i}(\bm{r},t),\label{E13c}\\
\vec{j}_{TV}^{\mu}(\bm{r},t)  &  =\sum_{i}\bar{\psi}_{i}(\bm{r},t)\vec{\tau
}\gamma^{\mu}\psi_{i}(\bm{r},t).
\end{align}
As in all relativistic mean field models, the \textit{no-sea}
approximation is used in the calculations of the nuclear densities by
summing only over the single-particle states with energies in the
Fermi sea. Vacuum polarization effects are not taken into account
explicitly but only in a global way by the correct choice of the Lagrangian
parameters. All interactions in the Lagrangian
(\ref{Lag-PC}) are then expressed in terms of the corresponding local
densities

Many effects, which go beyond mean field, seem to be neglected on the
classical level, such as Fock-terms, vacuum polarization, short range
Brueckner correlations etc. However, the coupling constants of the method are
adjusted to experimental data, which, of course, contain all these effects and
many more. Therefore these effects are not neglected. On the contrary, they
are taken into account in an effective way. This concept of RMF methods is
therefore equivalent to that of density functional theory.

The time-dependent variational principle
\begin{equation}
\delta\int\left\{  i\langle\Phi(t)|\frac{\partial}{dt}|\Phi(t)\rangle
-E[\hat{\rho}(t)]\right\}  \, dt=0 \label{TD_variation}%
\end{equation}
allows us to derive from the energy density functional $E[\hat{\rho}]$ an
equation of motion for the time-dependent relativistic single particle
density:
\begin{equation}
\label{density_matrix}\hat\rho(\mathbf{r},\mathbf{r^{\prime}},t)=\sum_{i}^{A}
|\psi_{i}(\mathbf{r},t)\rangle\langle\psi_{i}(\mathbf{r^{\prime}},t)|,
\end{equation}
which has the form
\begin{equation}
\label{TDRMF}i\partial_{t}\hat\rho(t)=[\hat{h}(\hat\rho(t)),\hat\rho(t)].
\end{equation}
The self energy, i.e. the single particle hamiltonian
$\hat{h}(\hat\rho(t))$ is obtained as the functional derivative of
the energy density functional with respect to the relativistic
density matrix:
\begin{equation}
\hat{h}=\frac{\delta E[\hat{\rho}]}{\delta\hat{\rho}}. \label{hph1}
\end{equation}%
This yields the Dirac hamiltonian:
\begin{equation}
\hat{h}=\bm{\alpha}[-i\bm{\nabla}-\bm{V}(\bm{r},t)]%
+V(\bm{r},t)+\beta(m+S(\bm{r},t))%
\label{Dirac-hamiltonian}
\end{equation}
with the self-consistent scalar and vector potentials
\begin{align}
S(\bm{r},t)  &  = \Sigma_{S}(\bm{r},t)+\vec{\tau}\cdot\vec{\Sigma}%
_{TS}(\bm{r},t),\\
V^{\mu}(\bm{r},t)  &  = \Sigma^{\mu}(\bm{r},t)+\vec{\tau}\cdot\vec{\Sigma
}^{\mu}_{T}(\bm{r},t).
\end{align}
The nucleon isoscalar-scalar, isovector-scalar, isoscalar-vector and
isovector-vector self-energies are density dependent and defined by the
following relations:
\begin{align}
\Sigma_{S}  &  = \alpha_{S}\rho_{S}+ \beta_{S}\rho_{S}^{2}+\gamma_{S}\rho
_{S}^{3}-\delta_{S}\Delta\rho_{S},\\
\vec{\Sigma}_{TS}  &  = \alpha_{TS}\rho_{TS}-\delta_{TS}\Delta\rho_{TS},\\
\Sigma^{\mu}  &  = \alpha_{V}\rho_{V}+\gamma_{V}\rho_{V}^{3}-\delta_{V}
\Delta\rho_{V} -eA^{\mu}\frac{1-\tau_{3}}{2},\\
\vec{\Sigma}^{\mu}_{T}  &  = \alpha_{TV}\rho_{TV}-\delta_{TV}\Delta\rho_{TV}.
\end{align}
Here we have neglected retardation effects, i.e. second derivatives with
respect to the time for the various densities.

In the static limit we have
\begin{equation}
\label{static}[\hat{h}(\hat\rho),\hat\rho]=0,
\end{equation}
thus the static density $\hat\rho_{0}$ is obtained from the solution of the
self-consistent Dirac equations upon all the nucleons with eigenvalues
$\varepsilon_{k}$ and eigenfunctions $\psi_{k}(r)$:
\begin{equation}
\label{Dirac}\hat{h}|\psi_{k}(\bm{r})\rangle=\varepsilon_{k}|\psi_{k}%
(\bm{r})\rangle.
\end{equation}
For spherical symmetry the spinors have the form:
\begin{equation}
|\mathcal{\psi}_{n\kappa\,m}(\bm{r})\rangle=\frac{1}{r}{\binom{f_{n\kappa
}(r)\mathcal{Y}_{\kappa\,m}(\Omega)}{ig_{n\kappa}(r)\mathcal{Y}_{\bar{\kappa
}\,m}(\Omega)}}. \label{wavefunctions}%
\end{equation}
The subscripts $n$, $\kappa$ and $m$ are principal and angular momentum
quantum numbers; $\kappa=\mp(j+\frac{1}{2})$ for $j=l\pm\frac{1}{2}$, where
$j$ and $l$ are the total and the orbital angular momenta of the nucleon. As
usual, $m$ is the $z$ component of the total angular momentum. The spherical
spinors $\mathcal{Y}_{\kappa\,m}(\Omega)$ are given in terms of spherical
harmonics $Y_{lm_{l}}(\Omega)$ and Pauli spinors $\chi_{m_{s}}$ as:
\begin{equation}
\mathcal{Y}_{\kappa\,m}(\Omega)=\sum_{m_{l}m_{s}}(\frac{1}{2}m_{s}%
lm_{l}|jm)Y_{lm_{l}}(\Omega)\chi_{m_{s}}, \label{harmonics}%
\end{equation}
while the functions $f_{i}(r)$ and $g_{i}(r)$ satisfy the static
radial Dirac equations:
\begin{equation}
\left(
\begin{array}
[c]{cc}%
V+S & -\partial_{r}+\frac{\kappa}{r}\\
\partial_{r}+\frac{\kappa}{r} & V-S-2m
\end{array}
\right)  \left(
\begin{array}
[c]{c}%
f_{i}(r)\\
g_{i}(r)
\end{array}
\right)  =\left(
\begin{array}
[c]{c}%
f_{i}(r)\\
g_{i}(r)
\end{array}
\right)  \varepsilon_{i}. \label{Dirac-radial}
\end{equation}

\begin{table}[t]
\centering
\renewcommand{\arraystretch}{1.5}%
\begin{tabular}
[c]{|l|r@{.}l|}\hline ~~Coupling const.~~ &
\multicolumn{2}{c|}{PC-F1}\\\hline
~~~~~~~~~$\alpha_{S}$~~~ & ~~-14 & 935894~~~\\
~~~~~~~~~$\delta_{S}$~~~ & ~~ -0 & 634576\\\hline
~~~~~~~~~$\alpha_{V}$~~~ & 10 & 098025\\
~~~~~~~~~$\delta_{V}$~~~ & -0 & 180746\\\hline
~~~~~~~~~$\alpha_{TS}$ & 0 & 0\\
~~~~~~~~~$\delta_{TS}$ & 0 & 0\\\hline
~~~~~~~~~$\alpha_{TV}$ & 1 & 350268\\
~~~~~~~~~$\delta_{TV}$ & -0 & 063680\\\hline\hline
~~~~~~~~~$\beta_{S}$ & 22 & 994736\\
~~~~~~~~~$\gamma_{S}$ & -66 & 769116\\\hline ~~~~~~~~~$\gamma_{V}$ &
-8 & 917323\\\hline
\end{tabular}
\caption{The coupling constants in the parameter set PC-F1 resulting
from the fitting procedure in Ref.~\cite{BMM.02}. The units are
[fm$^{-2}$] for the constants $\alpha$ of the quadratic terms,
[fm$^{-4}$] for the constants $\delta$ of the derivative terms,
[fm$^{-5}$] for the constants $\beta$ of the cubic terms, and
[fm$^{-8}$] for the constants $\gamma$ of the quartic terms
in the Lagrangian.}%
\label{tab1}%
\end{table}

The point coupling Lagrangian used in this work contains eleven
coupling constants. Based on an extensive multi parameter $\chi^{2}$
minimization procedure, B\"urvenich et al.~\cite{BMM.02} have
adjusted the parameter set $PC$-$F1$ to reproduce ground state
properties of infinite nuclear matter and spherical doubly closed
shell nuclei. This set is listed in Table~\ref{tab1} and it has
been tested in the calculation of many ground state properties of
spherical and deformed nuclei all over the periodic table. The
results are very well comparable with reasonable effective
meson-exchange interactions.

The nuclear ground state is defined as the equilibrium point of the functional
(\ref{Energy}), thus, is associated with the density which minimizes
$E_{\mathrm{RMF}}[\hat{\rho}]$. Furthermore, small oscillations around this
equilibrium point correspond to the vibrational nuclear states. They are
usually described within the harmonic approximation, that is, using linear
response theory. In nuclear physics, this is the so called Random Phase
Approximation (RPA) which has been already mentioned in our discussion and
will be described in more detail in the next section.

\section{Relativistic RPA formalism}

\label{CRPA}

Under the influence of an external field $F(\omega)$ oscillating with
the frequency $\omega$ the nucleus is excited. The cross section of
this process is proportional to the strength function:
\begin{align}
S(\omega)&=-\frac{1}{\pi}\operatorname{Im}%
\sum_{\alpha\beta\alpha^{\prime}\beta^{\prime}}%
F_{\alpha\beta^{\ast}}R_{\alpha\beta\alpha^{\prime}\beta^{\prime}}(\omega)%
F_{\alpha^{\prime}\beta^{\prime}}\nonumber\\
&:=-\frac{1}{\pi}\operatorname{Im}R_{FF}(\omega), \label{strenghtfunction}
\end{align}
where $F_{\alpha\beta}$ is the operator inducing the reaction and
$R_{\alpha\beta\gamma\delta}(\omega)$ is the response function which,
in an arbitrary representation indicated by the Greek indices
$\alpha,\beta,\ldots$ (e.g. the ($\bm{r},s)$-representation) is
defined as:
\begin{equation}
\label{full_response}
R_{\alpha\beta\alpha^{\prime}\beta^{\prime}}(\omega)%
=\sum\limits_{\nu}\left\{\frac{\langle0|a_\beta^+a _\alpha|\nu\rangle \langle\nu|a_{\alpha^{\prime}}^{+}a _{\beta^{\prime}}|0\rangle}{\omega-E_{\nu}+E_{0}+i\eta}\right. -\left.\frac{\langle\nu|a_\beta^+a _{\alpha}|0\rangle
\langle0|a_{\alpha^{\prime}}^{+}a _{\beta^{\prime}}|\nu\rangle}{\omega+E_{\nu}-E_{0}+i\eta}\right\}.
\end{equation}
The imaginary part $i\eta$ is infinitesimal and is introduced in
order to fulfill the proper boundary conditions and to prevent
$R(\omega)$ from diverging at $\omega=E_{\nu}-E_{0}$. We use here the
response derived from the retarded Green's functions as defined in
Ref.~\cite{FW.71}

In the independent particle model, $|0\rangle$ is the Slater
determinant of the ground state, formed by the self-consistent
solutions of the Dirac equation~(\ref{Dirac}) and
$|\nu\rangle=a_{p}^{+}a_{h}|0\rangle$ are $ph$-states, while $E_{0}$
and $E_{\nu}$ are the corresponding energies. In the basis
$|k\rangle$, where the single particle
hamiltonian~(\ref{Dirac-hamiltonian}) is diagonal we obtain the free
response function:
\begin{equation}
R_{klk^{\prime}l^{\prime}}^{\,0}(\omega)=\frac{n_{k}-n_{l}}{\omega
-\varepsilon_{k}+\varepsilon_{l}+i\eta}\delta_{kk^{\prime}}\delta_{ll^{\prime
}} \label{R0}%
\end{equation}
with the occupation factors:
\begin{equation}
n_{k}=\langle0|a_{k}^{+}a_{k} |0\rangle=\left\{
\begin{array}
[c]{ll}%
1 & \text{for hole states with }\varepsilon_{k}\leq\varepsilon_{F}\\
0 & \text{for particle states with }\varepsilon_{k}>\varepsilon_{F}%
\end{array}
\right.
\end{equation}

The full response of Eq.~(\ref{full_response}) contains the transition densities:
\begin{equation}
\rho_{\alpha\beta}^{\nu}=\langle0|a_\beta^{+}a _\alpha|\nu\rangle.
\label{transition_density}%
\end{equation}
They can be deduced from the time-dependent density matrix in
Eq.~(\ref{density_matrix}), which is derived from the variational
principle in Eq.~(\ref{TD_variation}).

In the small amplitude limit one uses the linear response
approximation to obtain the full response $R(\omega)$ of
Eq.~(\ref{full_response}) as the solution of the \textit{linearized
Bethe-Salpeter equation}:
\begin{equation}
\label{Bethe-Salpeter}
R_{\alpha\beta\alpha^\prime\beta^\prime} (\omega)=R_{\alpha\beta\alpha^\prime\beta^\prime}^{\,0}(\omega)
+\sum_{\gamma\delta\gamma^\prime\delta^\prime}\,\,R^0_{\alpha\beta\gamma\delta}(\omega)
V_{\gamma\delta\gamma^\prime\delta^\prime}^{\text{ph}} R_{\gamma^\prime\delta^\prime\alpha^\prime\beta^\prime}(\omega).
\end{equation}
The relativistic residual interaction is found as the second
derivative of the energy density functional~(\ref{Energy}) with
respect to the density matrix
\begin{equation}
V_{\alpha\beta\alpha^{\prime}\beta^{\prime}}^{\text{ph}}=\frac{\delta
^{2}E[\hat{\rho}]}{\delta\hat{\rho}_{\alpha\beta}\delta\hat{\rho}%
_{\alpha^{\prime}\beta^{\prime}}}.%
\label{energy_variation}%
\end{equation}
Once again, we have neglected retardation and this effective
interaction has to be calculated at the static density.

In a short hand notation the response equation~(\ref{Bethe-Salpeter})
has the formal solution
\begin{equation}
R (\omega)=(1-R ^{\,0}(\omega)V ^{\text{ph}})^{-1}R^{\,0}(\omega)
\label{formal-solution}%
\end{equation}
or introducing the inverse of $R^{\,0}$ we have%
\begin{equation}
R (\omega)=\frac{1}{R^{\,0}(\omega)^{-1}-V ^{\text{ph}}}
\label{RPA-inversion}%
\end{equation}

The evaluation of the strength function (\ref{strenghtfunction})
requires therefore three steps. The starting point is the calculation
of the free response function $R^{\,0}(\omega)$. In the next step one
determines the interaction  $V ^{\text{ph}}$ and finally one solves the response equation by the inversion
(\ref{formal-solution}). In details there are several methods to
proceed. In particular one can choose various basis sets to solve
these  equations.

a) As we have seen in Eq. (\ref{R0}) the free response has a
particularly simple form in the basis of Dirac spinors (\textit{Dirac
basis}) diagonalizing the self-consistent mean field
equation~(\ref{Dirac}). This is in particular simple for cases where
the Dirac equation is solved in a discrete basis, as for instance the
oscillator basis~\cite{GRT.90} or in a Saxon Woods
basis~\cite{ZMR.03} determined by the solution of the Dirac equation
in a box with finite size. However, the simplicity in the calculation
of $R^{\,0} (\omega)$ is compensated by the computational effort
required in the next steps. First we have to calculate a large number
of matrix elements for the interaction~(\ref{energy_variation}) in
the basis of the corresponding $ph$-states and in a second step the
matrix $(1-R^{\,0}(\omega)V^{\text{ph}})$ has to be inverted for each
value of the frequency $\omega$. In general the number  of single
particle states is rather large and this leads to a huge number of
$ph$-states, requiring considerable computational sources, not only
in memory but also in computer time. This is in particular a problem
in the case of  deformed nuclei. By this reason this method can only
be used successfully for light spherical nuclei, where the number  of
$ph$-states is limited.

b) The inversion is particular simple in the \textit{RPA-basis}.
Inserting expression~(\ref{R0}) into Eq.~(\ref{RPA-inversion}) we
find that the response function is equivalent to the resolvent of the
RPA matrix
\begin{equation}
R^{\,0}(\omega)^{-1}-V ^{\text{ph}}=\omega-\left(
\begin{array}
[c]{cc}%
A & B\\
-B^{\ast} & -A^{\ast}%
\end{array}
\right)  \label{RPA-matrix}%
\end{equation}
where%
\begin{align}
A_{php^{\prime}h^{\prime}} &  =(\varepsilon_{p}-\varepsilon_{h})\delta
_{pp^{\prime}}\delta_{hh^{\prime}}+V_{php^{\prime}h^{\prime}}^{\text{ph}%
},\text{\ }\\
B_{php^{\prime}h^{\prime}} &  =V_{phh^{\prime}p^{\prime}}^{\text{ph}}%
\end{align}
Of course, the calculation of this matrix requires the same numerical effort
as the evaluation of $V ^{\text{ph}}$ in the Dirac basis discussed above.
However there exist standard routines for the diagonalization of the
RPA-matrix
\begin{equation}
\left(
\begin{array}
[c]{cc}%
A & B\\
-B^{\ast} & -A^{\ast}%
\end{array}
\right)  \left(
\begin{array}
[c]{c}%
X\\
Y
\end{array}
\right)  _{\mu}=\left(
\begin{array}
[c]{c}%
X\\
Y
\end{array}
\right)  _{\mu}\Omega_{\mu}\label{RPA-diagonalization}%
\end{equation}
and this diagonalization has to be carried out only once, whereas the
inversion of the response equation has to be done for each value of
the frequency $\omega$. In the RPA-basis given by the eigenvectors
$|\mu\rangle$ the reduced response function defined in
Eq.~(\ref{reduced-response}) has a particular simple form
\begin{equation}
R_{cc^{\prime}}(\omega)={\sum\limits_{\mu>0}}\frac{\langle0|Q_{c}^{\dag}%
|\mu\rangle\langle\mu|Q_{c^{\prime}}|0\rangle}{\omega-\Omega_{\mu}+i\eta
}-\frac{\langle\mu|Q_{c}^{\dag}|0\rangle\langle0|Q_{c^{\prime}}|\mu\rangle
}{\omega+\Omega_{\mu}+i\eta}.
\label{Resp-RPA}%
\end{equation}
Using
\begin{equation}
\langle0|F|\mu\rangle=\sum\limits_{ph}F_{ph}(X_{ph}^{\mu}+Y_{ph}^{\mu})
\end{equation}
we find for $R_{FF}(\omega)$%
\begin{equation}
R_{FF}(\omega)={\sum\limits_{\mu>0}}\frac{|\langle0|F|\mu\rangle|^{2}}%
{\omega-\Omega_{\mu}+i\eta}-\frac{|\langle0|F|\mu\rangle|^{2}}{\omega
+\Omega_{\mu}+i\eta}%
\end{equation}
and for the strength function in Eq.~(\ref{strenghtfunction})%
\begin{align}
S(\omega+i\frac{\Delta}{2}) &  =-\frac{1}{\pi}\operatorname{Im}R_{FF}%
(\omega+i\frac{\Delta}{2})\label{E54}\\
&  ={\sum\limits_{\mu}}|\langle0|F|\mu\rangle|^{2}\frac{1}{2\pi}\frac{\Delta
}{(\omega-\Omega_{\mu})^{2}+\frac{1}{4}\Delta^{2}}\nonumber
\end{align}
Here $\Delta$ is a smearing parameter, which introduces a folding
with a Lorentzian and is introduced by numerical reasons.

c) In many cases the effective interaction
$V_{\alpha\beta\alpha^{\prime} \beta^{\prime}}^{\text{ph}}$
can formally be written as a sum of separable terms.
\begin{equation}
V_{\alpha\beta\alpha^{\prime}\beta^{\prime}}^{\text{ph}}=\sum_{c}
Q_{\alpha\beta}^{c}V_{c}^{\text{ph}}Q_{\alpha^{\prime}%
\beta^{\prime}}^{\dag\,c} \label{separable}%
\end{equation}
where $Q^{c}$ are single particle operators characterized by the
channel index $c.$ As discussed in  Appendix~\ref{AppA}, this is
particular the case for the effective interaction of the relativistic
point coupling model PC-F1 used in the present investigation. Working
in the channels given by these operators $Q^{c}$ the numerical effort
can be simplified considerably.

We insert the effective interaction (\ref{separable}) into the
Bethe-Salpeter equation (\ref{Bethe-Salpeter}) and introducing the
reduced response function:
\begin{equation}
R_{cc^{\prime}}(\omega)=\sum_{\alpha\beta\alpha^{\prime}\beta^{\prime}%
}\,\,Q_{\alpha\beta}^{c\dag}R_{\alpha\beta\alpha^{\prime}\beta^{\prime}%
}(\omega)Q_{\alpha^{\prime}\beta^{\prime}}^{c^{\prime}},
\label{reduced-response}%
\end{equation}
equation (\ref{Bethe-Salpeter}) turns into the reduced Bethe Salpeter
equation
\begin{equation}
R _{cc^{\prime}}(\omega)
=R_{cc^{\prime}}^{\,0}(\omega)%
+\sum_{c^{\prime\prime}}%
R_{cc^{\prime\prime}}^{\,0}(\omega)V_{c^{\prime\prime}}^{\text{ph}%
}R _{c^{\prime\prime}c^{\prime}}(\omega). \label{reduced-response-eq}%
\end{equation}
which has the same formal solution as given in Eq.
(\ref{formal-solution}). In all cases, where one has a continuous
channel index $c$, as for instance the radial coordinate $r$, this is
an integral equation. In Eq.  (\ref{reduced-response-eq}) the
interaction $V_{c }^{\text{ph}}$ is diagonal with respect to the
channel index $c$. This is not always the case. However, as we shall
see in Appendix~\ref{AppA}, the relativistic interaction PC-F1 can
be expressed to a large extent in this way. We have to allow only in
specific cases also for non-diagonal interactions
$V_{cc^{\prime}}^{\text{ph}}$, as for instance in the case of the
Coulomb force or in the case of derivative terms.  This is a rather
simple extension of the present method and therefore, for the sake of
simplicity, we will restrict ourselves in the following to an
interaction diagonal in the cannel index $c$. If the external
operator $F$ in Eq.  (\ref{strenghtfunction}) can be expressed by the
operators $Q_{c}$ as
\begin{equation}
F=\sum_{c}f_{c}Q_{c}%
\end{equation}
we finally obtain the strength function as%
\begin{equation}
S(\omega)=-\frac{1}{\pi}\operatorname{Im}R_{FF}=-\frac{1}{\pi}%
\operatorname{Im}\sum_{cc^{\prime}}f_{c}^{\ast}R_{cc^{\prime}} %
(\omega)f_{c^{\prime}} .%
\label{E55}%
\end{equation}
If $F$ cannot be expressed in terms of the operators $Q_{c}$ we
obtain $R_{FF}$ from the Bethe-Salpeter equation (\ref{Bethe-Salpeter}) as
\begin{equation}
R_{FF} (\omega)  =R_{FF}^{\,0}(\omega) +\sum_{cc^{\prime}}R_{F c}^{\,0}(\omega)V_{c}(1-R ^{\,0}
(\omega)V ^{\text{ph}})_{cc^{\prime}}^{-1}R_{c^{\prime}F}^{\,0}(\omega).
\end{equation}

\section{Treatment of the continuum.}

\label{continuum}

As we have briefly discussed earlier, a proper treatment of the
continuum is not possible by using a discrete basis, because one
needs a tremendously large number of $ph$-states to fill up the
continuum with. Instead, it can only be  properly taken into account
if one makes use of the more flexible linear response formalism in an
appropriate channel space.

Starting from Eq.~(\ref{reduced-response}) for the reduced response
function and using Eq.~(\ref{Resp-RPA}) we derive the following
expression for the reduced free response, which depends only on the
energy $\omega$ and the channel indices $c, c^{\prime}$:

\begin{equation}
R_{cc^{\prime}}^{0}(\omega)=%
\sum\limits_{ph}%
\frac{\langle h|Q_{c }^{+}|p\rangle%
\langle p|Q_{c^{\prime}}|h\rangle}%
{\omega-\varepsilon_{p}+\varepsilon_{h}}%
-\frac{\langle p|Q_{c}^{+}|h\rangle%
\langle h|Q_{c^{\prime}}|p\rangle}%
{\omega+\varepsilon_{p}-\varepsilon_{h}}%
\label{R0QQ}%
\end{equation}
where $h$ stands for occupied (hole) and $p$ for unoccupied
(particle) states. It is easy to show that the sum over $p$ can be
safely extended to run over the full space, since terms of the form
$\sum_{hh^{\prime}}$ vanish due to the cancellation of forward and
backward going parts. Using completeness we obtain:
\begin{eqnarray}
\label{QGQ}
R_{cc^{\prime}}^{0}(r,r';\omega) &=&
\sum_{h}\langle\,h|Q_{c}^{+} \frac{1}{\omega+\varepsilon_{h}-\hat{h}}Q_{c^{\prime}}-
Q_{c^{\prime}}\frac{1}{\omega-\varepsilon_{h}+\hat{h}}Q_{c}^{+}|h\rangle \nonumber \\
&=&
\sum_{h}\langle\,h|Q_{c}^{+}G(\omega+\varepsilon_{h})Q_{c^{\prime}}~+~Q_{c^{\prime}}G(-\omega+\varepsilon_{h})
Q_{c}^{+}|h\rangle.
\end{eqnarray}

Here, $\hat{h}$ is the Dirac hamiltonian (\ref{Dirac-hamiltonian}) and
$G(E)=1/(E-\hat{h})$ is the corresponding single particle Green's
function.

In this work we use relativistic zero range forces, thus it is
appropriate to work in coordinate space. The method described in the
following is a relativistic generalization of the method introduced
by Bertsch et al \cite{SB.75} for
non-relativistic zero range forces. In this case we solve the
response equation in $r$-space, which is considerably simpler than
the method introduced in Refs.~\cite{SRM.89} for finite range
forces.

In coordinate representation the indices $\alpha$,$\beta,\ldots$ in
Eq.~(\ref{full_response}) are abbreviations for the "coordinates"
$1=({\bm r}_1,d_1,s_1,t_1)$, where $s$ is the spin, $t$ the isospin
coordinate, and $d=1,2$ labels large and small components. Starting
from the energy density functional (\ref{Energy}) we find the
effective interaction in Eq.~(\ref{energy_variation}) to be of the
form (\ref{separable}):
\begin{equation}
V^{\text{ph}}(1,2)~=~%
{\displaystyle\sum\limits_{c}}
{\displaystyle\int\limits_{0}^{\infty}}
dr~Q_{c}^{(1)}(r)~\upsilon_{c}(r)~Q_{c}^{\dag(2)}(r)
\end{equation}
with the local channel operators $Q_{c}({r)}$ defined by
\begin{equation}
Q_{c}^{(1)}(r)=~\frac{\delta({r}-{r}_1)}{rr_{1}}\gamma^{(1)}_{D}
\left[ \sigma^{(1)}_{S}Y_{L} (\Omega_{1})\right]_{J} \tau^{(1)}_{T}
\label{channel}%
\end{equation}
where we distinguish the "coordinates" abbreviated by the upper index~(1) and the channel index ($r,c$) used in Eq.~(\ref{separable}). Due to this r-dependance, the dimension of the matrix $R^{0}_{cc^{\prime}}(r,r^{\prime};\omega)$ in the numerical applications will be the number of r-mesh points times eight, which represents the number of the covariant channels c, given in Table~\ref{tab6} of the Appendix~\ref{AppA}. This implies that all scalar, longitudinal, and transverse modes (isoscalar and isovector) are fully included and mixed by the matrix inversion of Eq.(\ref{formal-solution}).

This channel index has now a continuous part
given by the radial coordinate $r$ and a discrete part characterized
by the quantum numbers $c=(D,S,L,T)$ where the Dirac index $D$ runs
over three 2$\times$2 matrices $\gamma_D=$ $\gamma_{0},1,\gamma_{5}$
defined in Eq. (\ref{gamma-2}), $S=0,1$ is the spin, $L$ the orbital
angular momentum and $T=0,1$ the isospin. Further details are given
in Appendix~\ref{AppA}.

Inserting the channel operators (\ref{channel}) into
Eqs.~(\ref{reduced-response}) and (\ref{QGQ}) we obtain the reduced
free response function:
\begin{eqnarray}
\mathcal{R}_{cc^{\prime}}^{\,0}(r,r^{\prime};\omega)&=&{\sum\limits_{h\kappa}%
}\left\{  Q_{\kappa h}^{\ast c}Q_{\kappa h}^{c^{\prime}}\,\langle
h(r)|\gamma_{D}^{+}G_{\kappa} (r,r^{\prime};\omega+\varepsilon_{h}%
)\gamma_{D^{\prime}} |h(r^{\prime})\rangle\,\right.\\
&+&\left.Q_{h\kappa}^{\ast
c}Q_{h\kappa}^{c^{\prime}}\langle h(r^{\prime})|\gamma_{D^{\prime}}^{{}%
}G_{\kappa} (r^{\prime},r;-\omega+\varepsilon_{h})\gamma_{D }%
^{+}|h(r)\rangle\right\}.\nonumber
\label{E59}
\end{eqnarray}
The sum runs over all the occupied states (hole) states $h$ with the
2-dimensional radial Dirac spinor $\langle
h(r)|=(f^\ast_{h}(r)~g^\ast_{h}(r))$ in Eq. (\ref{Dirac-radial}) and
over all the quantum numbers $\kappa=(lj)$ compatible with the
selection rules in the reduced angular and isospin matrix elements
\begin{equation}
Q_{h\kappa}^{c}:{=~e}_{T_{c}}\langle\kappa_{h}||\left[  \sigma_{S_{c}}Y_{L_{c}} \right]  _{J}||\kappa\rangle,
\end{equation}
where $e_{T_{c}}=1$ in the isoscalar channel ($T_{c}=0$) and
$e_{T_{c}}=\pm1$ (for protons or neutrons) in the isovector channel
($T_{c}=1$). The reduced matrix elements of the operator $\left[
\sigma_{S_{c}} Y_{L_{c}}^{{} }\right]  _{J}$ contain integrations
over the orientation angles $\Omega$ and sums over the spin indices.
The matrix elements of the form $\langle
h|\gamma_{D}G(E)\gamma_{D^{\prime}}|h\rangle$ depend on $r$ and
$r^{\prime}$ and are  obtained by summing over the Dirac indices
$d=1,2$ for large and small components.

The Green's function $G_{\kappa} (r,r^{\prime},E)$ describes the
propagation of a particle with the energy $E$ and the quantum numbers
$\kappa$ from $r$ to $r^{\prime}$. It can either be calculated by
\textit{spectral} or \textit{non-spectral} methods. In the
\textit{spectral }representation~\cite{BT.75} it is obtained as a
discrete sum
\begin{equation}
G_{\kappa} (r,r^{\prime};E)=\sum_{n}\frac{|n(r)\rangle\langle
n(r^{\prime})|}{E-\varepsilon_{n}}. \label{E49}%
\end{equation}
over a complete set of eigenstates $|n(r)\rangle$ of the radial Dirac
equation~(\ref{Dirac-radial}) with the quantum number $\kappa$ using
box boundary conditions (or an oscillator expansion). In this case
the continuum is discretized, in correspondence to the bound states
inside the potential. In principle, the radial quantum number $n$
runs over the whole single particle basis characterized by the
angular quantum number $\kappa$, but one can show that this is
identical to summing only over the unoccupied states, since the
hole-hole pairs in Eq.~(\ref{E59}) are not contributing, due to the
cancellation between forward and backward going part. Furthermore,
because of the no-sea approximation the states in the Dirac sea are
empty and therefore the sum over $n$ in Eq. (\ref{E49}) has also to
be extended over the negative energy states. This corresponds to the
sum over the $ah$-components discussed in the introduction. In
practical applications one has to restrict this infinite set by a
finite sum introducing an upper limit
$\epsilon_p-\epsilon_h<E^{ph}_{cut}$ in energy for the particle
states $p$ above the Fermi surface and a lower limit
$\epsilon_a-\epsilon_h > -E^{ah}_{cut}$ for the negative energy
solutions $a$ is introduced in order to make the - otherwise infinite
- sum, tractable. This leads to a discretized  spectrum.

In the spectral representation the response function
$\mathcal{R}^{\,0}(\omega)$ has poles at the $ph$-energies
$\omega=\pm(\varepsilon_{p}-\varepsilon_{h})$ and the full response
function $\mathcal{R}(\omega)$ has  poles at the eigenenergies
$\Omega_{\mu}$ of the RPA-equation (\ref{RPA-diagonalization}) in the
same restricted space. For real frequencies $\omega$ it is purely
real, and therefore the strength function vanishes everywhere apart
from these poles. For complex energies $\omega+i\Delta/2$, however,
these poles are shifted from the real axis and one  obtains a
continuous spectrum, with the phenomenological width $\Delta$. This
procedure yields identical results as the diagonalization of the
RPA-matrix in (\ref{RPA-diagonalization}) along with a subsequent
folding with a Lorentzian as  discussed in Eq. (\ref{E54}).

In the non-spectral or continuum approach~\cite{SB.75} the single
particle Green's function is constructed at each energy from two
linearly independent solutions of the Schroedinger equation with
different boundary conditions at $r=0$  and at $r\rightarrow\infty$.
In the relativistic case the Dirac-equation in $r$-space depending on
the quantum number  $\kappa$ is a two-dimensional equation and
therefore the corresponding single particle Green's function is a
2$\times2$ matrix. Using the bracket notation of Dirac for the
2-dimensional spinors we can write \cite{Tam.92}:
\begin{equation}
G_{\kappa} (r,r^{\prime};E)=\left\{
\begin{array}
[c]{cc}%
|w _{\kappa}(r)\rangle\langle u^\ast_{\kappa}(r^{\prime})| & \text{for}%
\,\,r>r^{\prime}\\
|u _{\kappa}(r)\rangle\langle w^\ast_{\kappa}(r^{\prime})| & \text{for}%
\,\,r<r^{\prime}%
\end{array}
\right.  \label{continuum-greens}%
\end{equation}
where $u(r)$ and $w(r)$ are two independent Dirac spinors~\cite{Tam.92}:
\begin{equation}
|u_{\kappa\,} (r)\rangle={\binom{f_{u}(r)}{g_{u}(r)},}\text{
\ \ \ \ \ \ \ \ }|w_{\kappa\,} (r)\rangle={\binom{f_{w}(r)}{g_{w}(r)}}
\label{scat_spinors}%
\end{equation}
normalized in such a way that the Wronskian
\begin{equation}
W=f_{w}(r)g_{u}(r)-g_{w}(r)f_{u}(r),
\end{equation}
which is independent of $r$, is normalized to unity. The solution
$u_{\kappa }(r)$ is regular at the origin and the solution
$w_{\kappa}(r)$ fulfills outgoing wave boundary conditions
\cite{Gre.90}. Further details are given in Appendix \ref{AppB}.

Provided that the free response function
$\mathcal{R}^{0}_{c,c^{\prime}}(r,r^{\prime};\omega)$ has been
properly derived, we are able to solve the reduced Bethe-Salpeter
equation~(\ref{reduced-response-eq})
\begin{eqnarray}
\mathcal{R}_{c,c^{\prime}}(r,r^{\prime};\omega)&=&
\mathcal{R}_{c,c^{\prime}}^{\,0}(r,r^{\prime};\omega)\\
&+&\sum_{c^{\prime\prime}}\int_{\,0}^{\infty}dr^{\prime\prime}\,
\mathcal{R}_{c,c^{\prime\prime}}^{0}(r,r^{\prime\prime};
\omega)\frac{\upsilon_{c^{\prime\prime}}(r^{\prime\prime})}{r^{\prime
\prime\,\,2}}\mathcal{R}_{c^{\prime\prime},c^{\prime}}(r^{\prime\prime
},r^{\prime};\omega).
\nonumber%
\label{E65}%
\end{eqnarray}

where the index $c^{\prime\prime}$ runs over the various discrete
channels given in Table~\ref{tab6}. Finally the strength function is
obtained as:
\begin{align}
\label{E68}
S(\omega)&  =-\frac{1}{\pi}\operatorname{Im}\mathcal{R}_{FF}\nonumber\\
&  =-\frac{1}{\pi}\operatorname{Im}%
{\displaystyle\iint\limits_{\,0}^{\infty}}
drdr^{\prime}F_{c}^{\ast}(r)\mathcal{R}_{cc^{\prime}} (r,r^{\prime}%
;\omega)F_{c^{\prime}} (r^{\prime}).
\end{align}
The sum rules are defined as moments of the strength function $S(\omega)$:
\begin{equation}
m_{k}=\int_{0}^{\infty}\omega^{k}S(\omega)\,d\omega.
\label{moments}%
\end{equation}
They are helpful to characterize the spectral distribution of the oscillator
strength. In particular they allow us to define the centroid energy by the
ratio%
\begin{equation}
E_{c}=\frac{m_{1}}{m_{0}}.%
\label{centroid-energy}%
\end{equation}
This quantity can be compared directly with experimental values. Of
course, in most experiments only a restricted energy range is
accessible and therefore one also has to restrict the integration in
Eq.~(\ref{moments}) to the same energy window.

Other important quantities are transition densities in various channels $c$
with respect to the operator $F$
\begin{equation}
\delta\rho_{c}(r;\omega)=\sum_{c^{\prime}}\int_{o}^{\infty}dr^{\prime}\mathcal{R}_{cc^{\prime}%
} (r,r^{\prime};\omega)F_{c^{\prime}} (r^{\prime}) \label{traden}%
\end{equation}
as for instance the neutron and proton transition densities:
\begin{equation}
\delta\rho(r)_{n,p}=\delta\rho_{T=0}(r;\omega)\pm\delta\rho_{T=1}(r;\omega)
\label{traden-pn}%
\end{equation}


\section{Applications}

\label{MGR}

In the previous section we briefly described how conventional RPA
methods treat the continuum part of the spectrum  through the
introduction of a potential "wall" far from the nucleus. In the
credit side of this approach, general properties of collective
excitations can be very well reproduced, either by using finite range
or point coupling interactions (Nik{\v{s}}i{\'{c}}
et.al.~\cite{NVR.05}). Since CRPA can treat the coupling to the
continuum exactly, it is of interest to see how well this model does
in reproducing the properties of excited state in finite nuclei, in
particular the giant resonances.

The most prominent resonances are the Isoscalar Monopole Resonance
(ISGMR), which is a breathing of the nucleus as a whole, the
Isovector Dipole Resonance (IVGDR) which corresponds to a collective
excitation of the proton against the neutron density, and Isoscalar
Quadrupole Resonance (ISGQR).  In addition we have the Isoscalar
Dipole Resonance (ISGDR) revealing the spurious state corresponding
to a translational motion of the nucleus. These modes show up in an
energy range of $10-30$ MeV and they exhaust a major portion of the
corresponding sum rules. In the next sections we
investigate the ISGMR, the IVGDR and the ISGDR in more detail.

\subsection*{Numerical details}

In the following, we perform several calculations using the
relativistic continuum RPA approach in $r$-space with Point Coupling
forces~\cite{BMM.02}. We select the doubly magic nuclei $^{16}$O,
$^{40}$Ca, $^{132}$Sn and $^{208}$Pb to investigate how the
collective excitation phenomena depend on an exact coupling to the
continuum.

In a first step, the ground state of the nucleus is determined by
solving the self-consistent RMF equations (\ref{Dirac-radial}) for
the parameter set PC-F1 given in Table~\ref{tab1}. The method we are
using is a fourth order Runge-Kutta in $r$-space (Dirac-mesh) where
nucleons move in a spherical box with radius $R_{D}=15$ fm and with a
mesh size $d_{D}=0.05$ fm.

Using the single particle wave functions and the corresponding
energies of this static solution, we determine the free response
$\mathcal{R}^{\,0}$ of Eq. (\ref{E59}) in the same box radius but
using a wider mesh in $r$-space (response-mesh). The size $d_{R}$ of
this mesh depends on the excitation mode; for the monopole modes we
use $d_{R}=0.15$ fm, while for the dipole a larger interval
$d_{R}=0.30$ fm is sufficient. Then we solve the Bethe salpeter
equation (\ref{E65}) to get the strength distribution $S(\omega)$.

At the same time, we perform similar calculations using the discrete
RPA approach, where the continuum is not treated exactly, aiming of
course to a more precise comparison with the CRPA results. For those
calculations, an energy cut-off is necessary, so that a feasible
diagonalization is achieved. In particular, we have used an energy
cut-off  $|\epsilon_{p}-\epsilon_{h}|< E^{ph}_{cut}= 300$ MeV for the
configurations with particles above the Fermi sea and $|\epsilon_{a}
-\epsilon_{h}|<E^{ah}_{cut}=1500$ MeV for configurations with
anti-particles in the Dirac sea.

\subsection*{Isoscalar Giant Monopole Resonances}

Results for the isoscalar monopole strength distribution are attainable, once the corresponding external field
\begin{equation}
F_{L=0}^{T=0}=\sum_{i}^{A}r^{2}_{i}
\end{equation}
is used. In this case, the classical energy weighted sum rule $m_{1}(E0)$ becomes:
\begin{equation}
m_{1}(E0)=\frac{1}{2}\langle[F,[T,F]]\rangle=\frac{\hbar^{2}}{2m}\langle
\nabla^{2} F\rangle=\frac{2\hbar^{2}}{m}\langle\,r^{2}\rangle.
\end{equation}
The doubly magic spherical nucleus $^{208}$Pb is a particularly good
example in perform our calculations, since it has been used in the
literature to test numerous nuclear structure models in the past, in
particular applications of the random phase approximation
\cite{RSp.74,Pie.00,Pie.01,CG.04}.

\begin{figure}[!t]
\centering
\includegraphics[width=350pt]{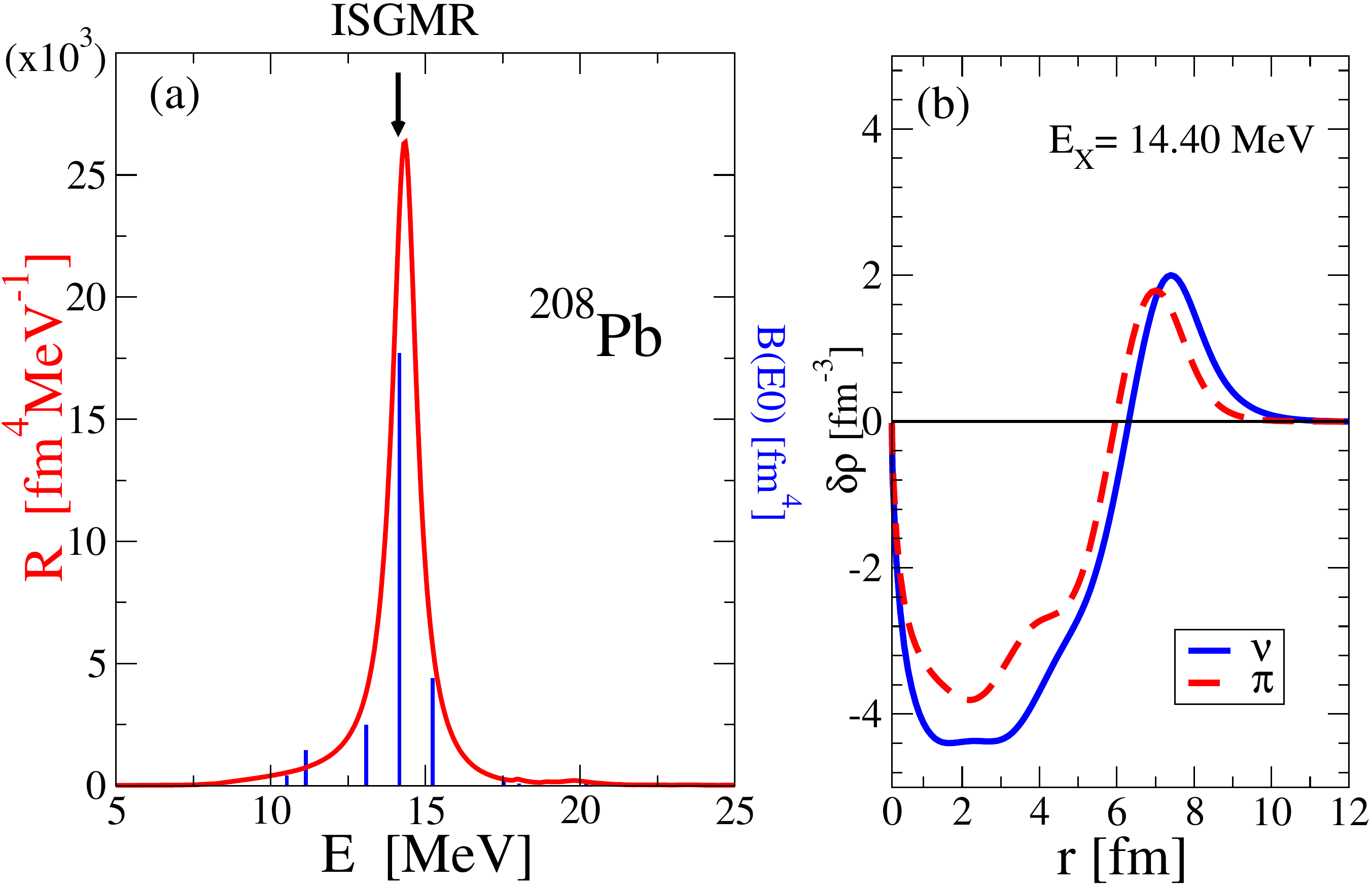}%
\caption{(Color online) (a) The isoscalar monopole spectrum in
$^{208}$Pb, calculated with the parameter set PC-F1. The red curve
corresponds to the strength distribution (units on the l.h.s.)
obtained by a non-spectral representation without smearing
($\Delta=0$), the blue lines give the discrete B(E0)-values (units on
the r.h.s.) obtained by the spectral representation with the same
force. The black arrow indicates the experimental centroid energy of
the resonance~\cite{YLC.04a}. (b) the neutron and proton transition
densities at the peak with the energy $E=14.40$
MeV.}%
\label{fig1}%
\end{figure}

In Fig.~\ref{fig1} we show the ISGMR strength distribution obtained
by continuum RPA (full red line) and compare it with the discrete
B(E0) values (blue) obtained by the spectral representation of the
response function for the same parameter set PC-F1~\cite{BMM.02}.

Using the CRPA approach, we find for the calculated centroid energy
defined in Eq. (\ref{centroid-energy}) that $m_{1}/m_{0}=14.40\,$
MeV, which is rather close to the result $m_{1}/m_{0}=14.17$ MeV
deduced from discrete RPA~calculations as well as to the experimental
value $m_{1}/m_{0}=13.96\pm0.2$~MeV \cite{YLC.04a}.

In those two methods, no additional smearing $\Delta=0$ has been
used. This means that the observed width of the continuum RPA
strength corresponds entirely to the escape width which in the Pb
region is very small, due to the relatively high Coulomb and
centrifugal barriers in this heavy nucleus. In contrast, discrete RPA
provides no width at all. Otherwise, the agreement of these two
methods in this nucleus is excellent.

In the panel (b) of Fig.~\ref{fig1}, we give the neutron and proton
transition densities at the peak energy, as it is calculated in
Eq.~(\ref{traden-pn}). They emphasize the collective character of the
isoscalar breathing mode extended over the entire interior of the
nucleus with neutrons and protons always in phase.

In addition, the energy weighted sum rule obtained in CRPA using Eq.
(\ref{moments}) is $m_{1} (E0)=5.448\cdot10^{5}$
[MeV$\cdot$fm$^{4}$]. This result is in excellent agreement with the
DRPA calculation $m_{1}(E0)=5.446\cdot10^{5}$ [MeV$\cdot$fm$^{4}$] as
well as the classical value $m_{1}(E0)=4A\hbar/2m\langle
r^{2}\rangle=5.453\cdot10^{5}$ [MeV$\cdot$fm$^{4}$]. This shows that
the results obtained in the literature by relativistic RPA
calculations using the spectral method  are very reliable for such
heavy nuclei \cite{VWR.00,RMG.01}.

\begin{figure}[t]
\centering
\includegraphics[width=350pt]{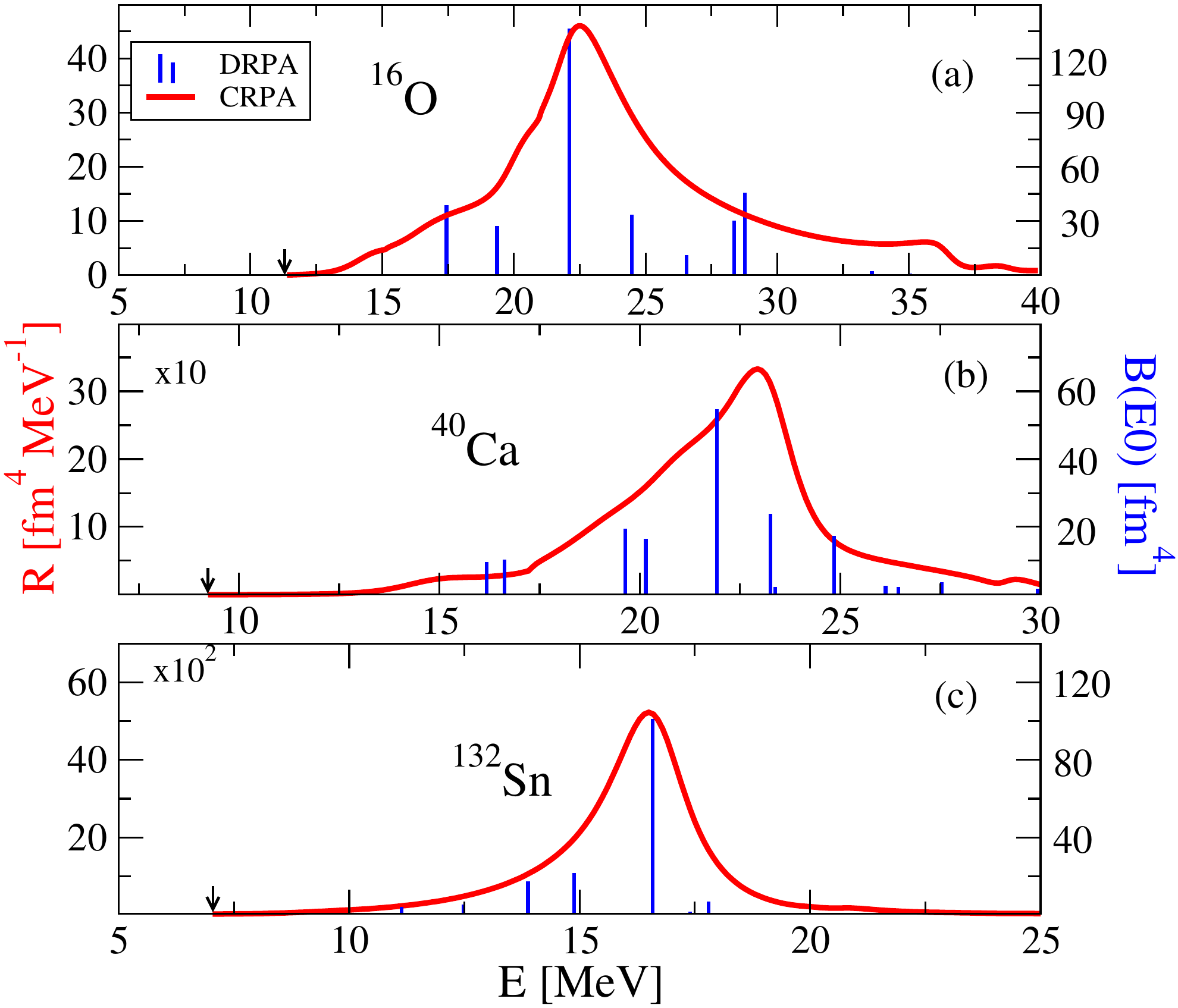}\caption{(Color online) The isoscalar
monopole strength distribution for doubly magic nuclei (a) in
$^{16}$O, (b) in $^{40}$Ca,
and (c) in $^{132}$Sn. Details are the same as in the panel (a)
of Fig. \ref{fig1}.}%
\label{fig2}%
\end{figure}

\begin{figure}[t]
\centering
\includegraphics[width=220pt]{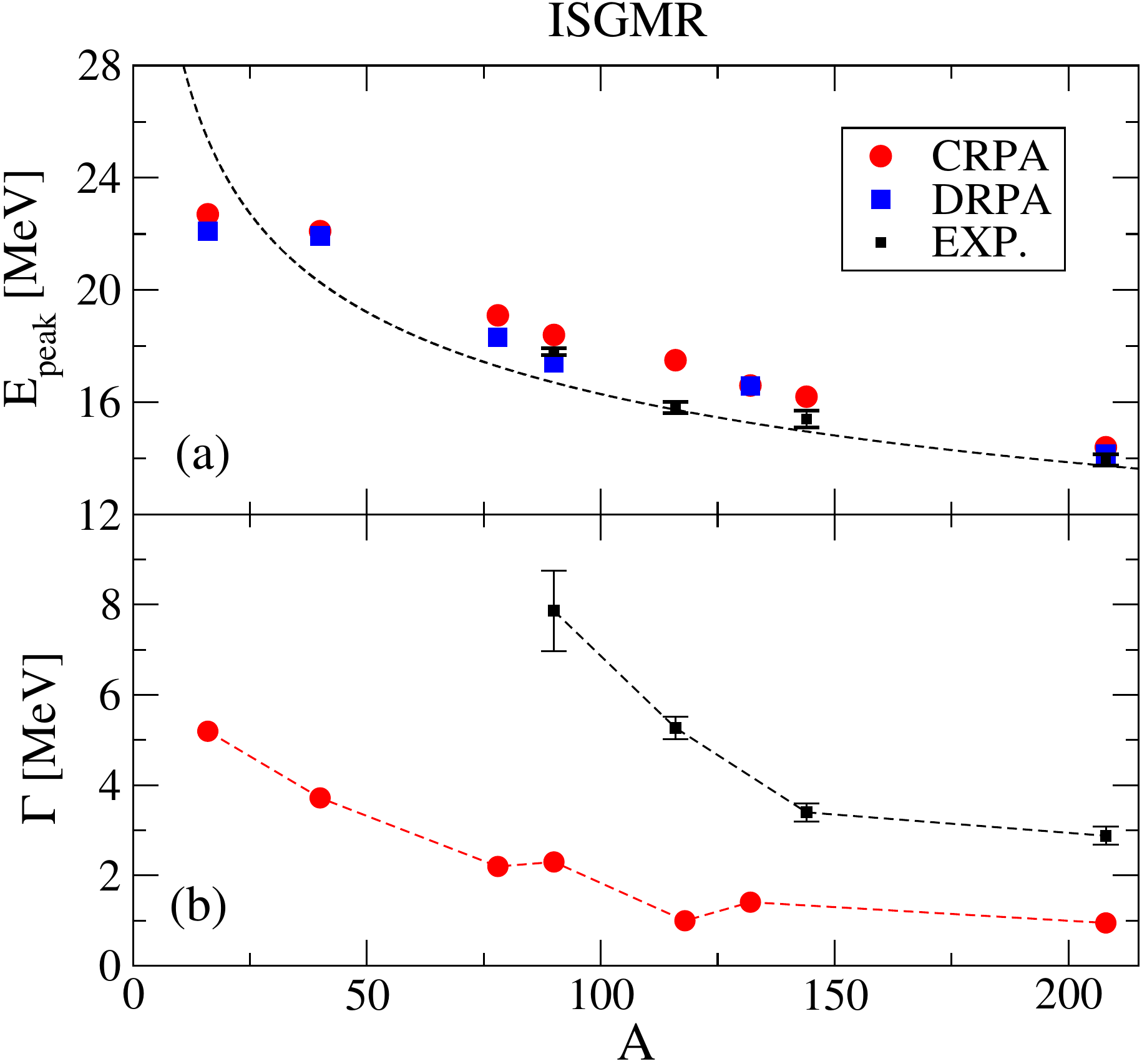}
\caption{(Color online) (a) The ISGMR centroid energies as a function
of the mass number, (b) The experimental and theoretical width of the
ISGMR as a function of the mass number. Details are given in the text.}%
\label{fig3}%
\end{figure}

In Fig.~\ref{fig2} we show the E0 strength distributions for the
lighter doubly magic nuclei $^{16}$O,  $^{40}$Ca, and $^{132}$Sn.  As
in Fig.~\ref{fig1}, the smearing parameter $\Delta$ is zero, but now
the escape width is considerably larger for these nuclei.
Fig.~\ref{fig3} summarizes the results for the isoscalar monopole
strength  distributions as a function of the mass number $A$. In
panel (a), we plot the centroid energies of both continuum RPA (red
dots) and discrete RPA (blue dots), together with the experimental
centroid energies taken from Ref.~\cite{YLC.04a}. We also show the
phenomenological $A$-dependence  $\bar{E}_{1^{-}}\approx
31.2\,A^{-1/3}+20.6\,A^{-1/6}$ by the dashed line. It becomes clear
that CRPA can successfully reproduce collective excitations over the
known range of nuclei.

In panel (b) of Fig.~\ref{fig3} we show the escape width
$\Gamma^{\uparrow}$ of E0 resonances. The red values correspond to
the full width half maximum (FWHM) of the peak, using continuum RPA ,
while the experimental values are indicated in black. The evident
disagreement is not surprising, if we consider that only
$1p1h$-configurations are taken into account, i.e. the major part of
the width resulting from the coupling to more complicated
configurations such as $2p2h$ etc. is not described well in this
simple RPA approach. It has been shown in recent investigations of
the coupling to complex configurations within the framework of the
relativistic time-blocking approximation (RTBA) or the
relativistic quasiparticle-time-blocking approximation
(RQTBA)~\cite{LRT.08} that such couplings can be taken into account
successfully in a fully consistent way starting from one density
functional $E[\rho]$. So far, relativistic investigations of this
type have been carried out with discrete methods. At present,
investigations in this direction including the continuum properly go
beyond the scope of this paper.

\subsection*{Isovector Giant Dipole Resonances}
\label{IVGD}

Isovector Giant Dipole resonance is the most well studied collective
excitation and the first to be observed experimentally~\cite{BK.47}.
An external electromagnetic field of the form:
\begin{equation}
F_{L=1}^{T=1}=\frac{N}{A}\sum_{p=1}^{Z}r_{p}Y_{1M}(\Omega_{p})-
\frac{Z}{A}\sum_{n=1}^{N}r_{n}Y_{1M}(\Omega_{n})
\end{equation}
causes protons and neutrons to oscillate in opposite  phases to each
other and this leads to a pronounced peak in the photoabsorption
cross section. This mode has been well studied in many
nuclei~\cite{Speth.91}.

With the increasing number of experiments in systems far from
stability and systems with large neutron excess, one has been able to
observe also low-lying E1 strength in the area of the neutron
emission threshold. It is called Pygmy Dipole Resonance PDR and can
be interpreted as a collective mode with dipole character where the
neutron skin oscillates against an isospin saturated proton-neutron
core.  This mode has first been predicted in phenomenological models
~\cite{MDB.71} exhausting several percent of the electric
dipole sum rule. In recent years, it has been intensively
investigated both on the experimental side by the Darmstadt
group~\cite{RHK.02,ZBH.05} as well as on the theoretical side, using
discrete relativistic RPA calculations based on
NL3~\cite{VPR.01a}.

\begin{figure}[!t]
\centering
\includegraphics[width=300pt]{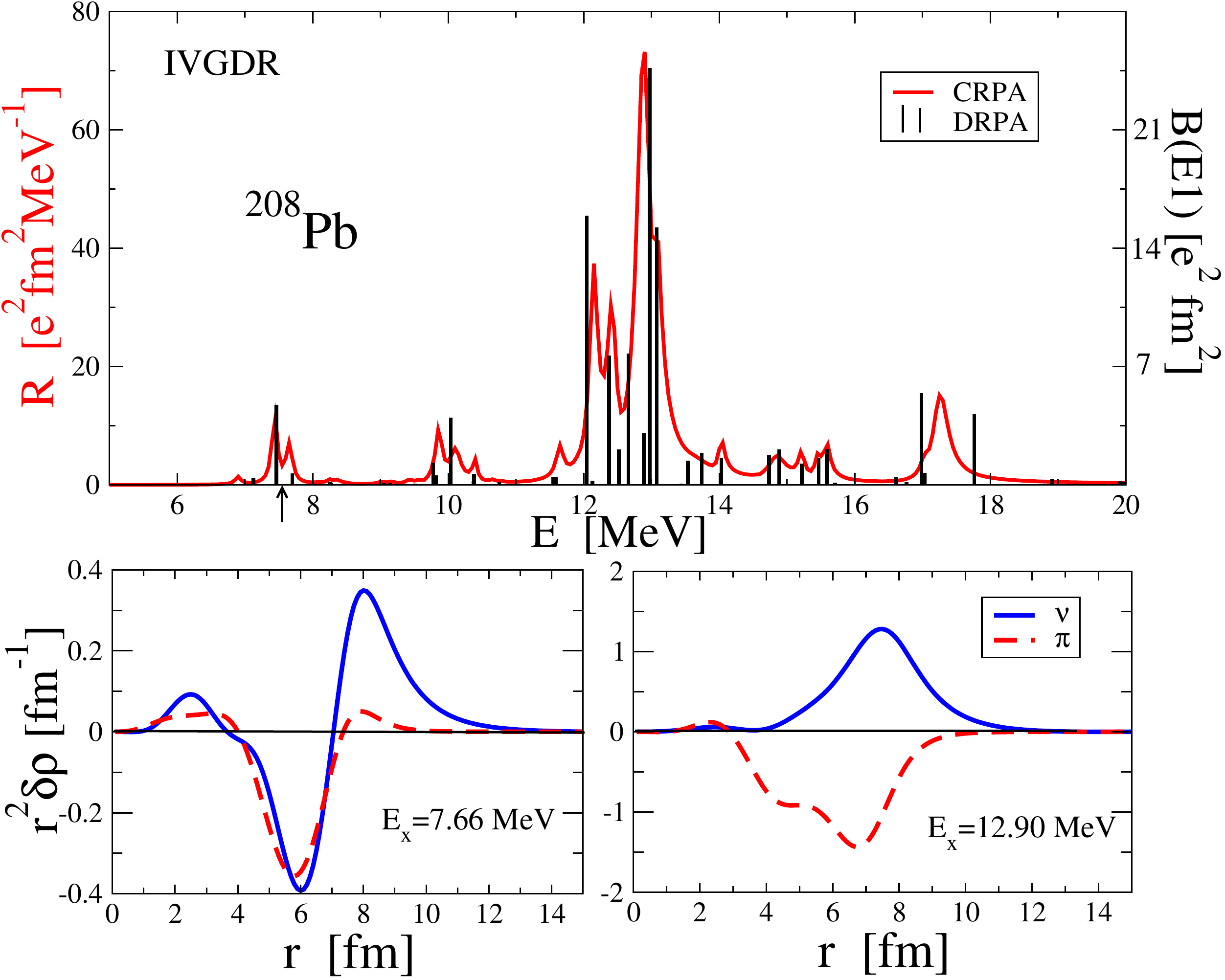}%
\caption{(Color online) (a) The isovector dipole strength
distribution in $^{208}$Pb. Details are essentially the same as in
the panel (a) of Fig. \ref{fig1}. However, in order to distinguish
the continuum (red curve) and the discrete (blue lines) calculations
we have used here a small smearing parameter $\Delta=10$ keV in the
continuum calculation. The black arrow indicates the theoretical
neutron emission threshold. (b) transition densities for neutrons and
(c) for protons
at the energy of the PDR (left) and at the GDR (right).}%
\label{fig4}%
\end{figure}

In Fig.~\ref{fig4} we show in panel (a) the results of the isovector
dipole strength E1 in the nucleus $^{208}$Pb using the CRPA approach.
The centroid energy at $13.32$ MeV is in excellent agreement with the
experimental excitation energy $E=13.3$~MeV \cite{RBK.93}. The energy
weighted sum rule~(\ref{moments}) is found as $m_{1}(E1)=916.28$
[MeV$\cdot$fm$^{2}$]. This result is in agreement with the DRPA
calculation, where we obtain $m_{1}(E1)=943.32$ [MeV$\cdot$fm$^{2}$]
and as usual somewhat (23.8 \%) larger than the classical
Thomas-Reiche-Kuhn sum
rule%
\begin{equation}
m_{\rm TRK}= \frac{9}{4\pi}\frac{\hbar^2}{2m}\frac{NZ}{A}=740.13~
[{\rm MeV}\cdot{\rm fm}^2].%
\label{TRK}
\end{equation}
In addition to the giant dipole resonance a smaller peak appears at
the energy region of the neutron emission threshold around
$E\sim7.5$~MeV, that corresponds to the pygmy resonance.

In panel (b) of Fig.~\ref{fig4} we give the transition densities
associated the low-lying peak at $E=7.66$ MeV and the GDR peak at
$E=12.9$ MeV. The higher peak has clearly an isovector character,
since the neutrons are oscillating against the protons over a large
radial range centered at the surface. The lower peak shows an
isoscalar core, where neutrons and protons oscillate in phase and a
pure neutron skin moving against the $T=0$ core. This is the typical
behavior of the pygmy mode.

\begin{figure}[!t]
\centering
\includegraphics[width=300pt]{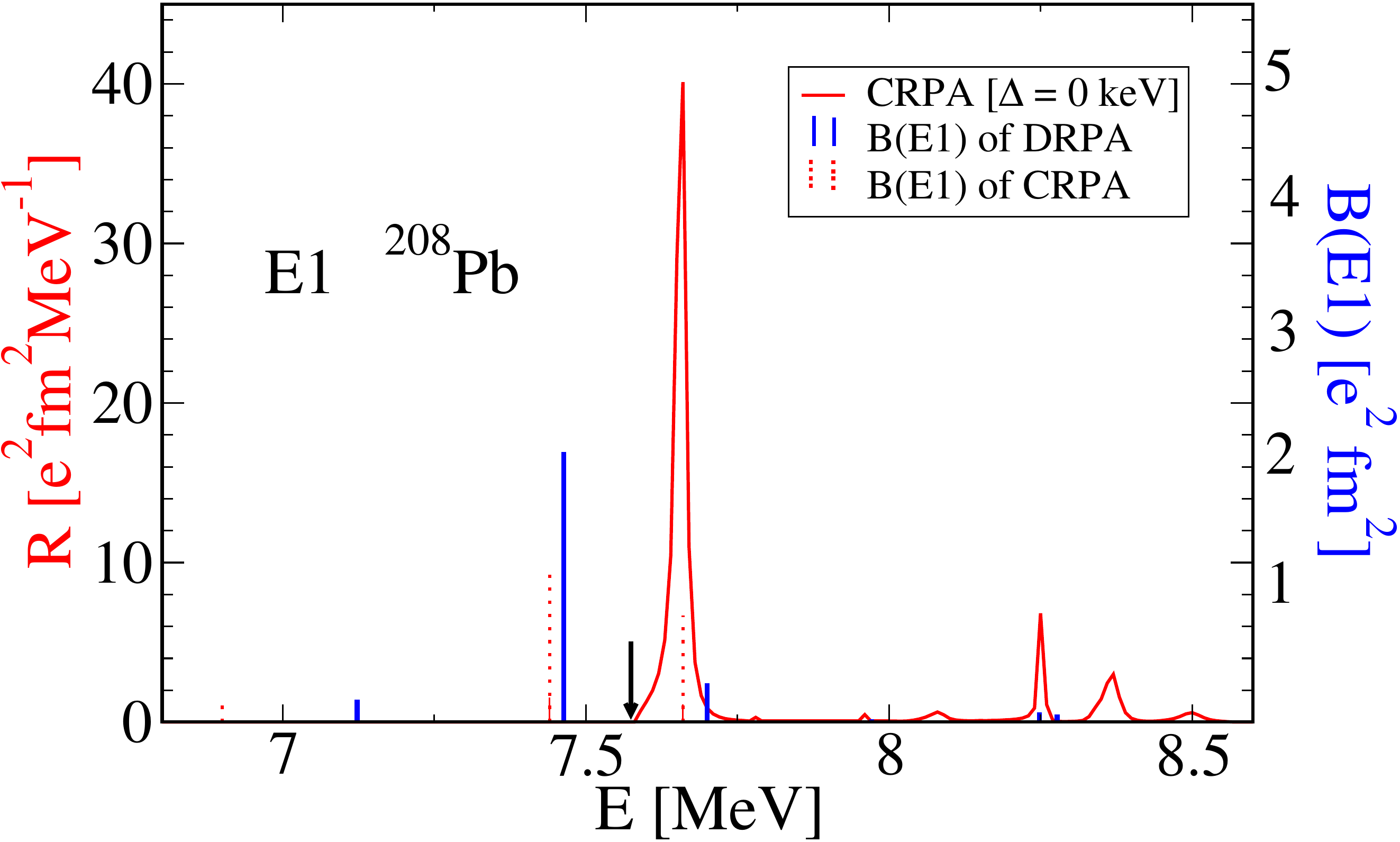} \caption{(Color online) The E1 pygmy
resonance (PDR) in the nucleus $^{208}$Pb. The black arrow indicates the theoretical neutron emission threshold at
$7.58$~MeV. The red dashed lines are obtained by CRPA calculations below the threshold.}%
\label{fig5}%
\end{figure}

\begin{table}[!b]
\centering
\renewcommand{\arraystretch}{1.5}%
\begin{tabular}
[c]{|c|r@{.}l|r@{.}l|r@{.}l|r@{.}l|}\hline
~~ & \multicolumn{4}{c|}{CRPA} & \multicolumn{4}{c|}{DRPA}\\\hline
~~No.~~ & \multicolumn{2}{c|}{~E~} & \multicolumn{2}{c|}{~B(E1)~} &
\multicolumn{2}{c|}{~E~} & \multicolumn{2}{c|}{~B(E1)~}\\\hline
~1~ & ~~~6 & 90~~ & ~~0 & 19~~ & ~~7 & 12~~ & ~~0 & 23~~\\
~2~ & ~~~7 & 44~~ & ~~1 & 45~~ & ~~7 & 46~~ & ~~2 & 82~~\\
~3~ & ~~~\textit{7} & \textit{66}~~ & ~~\textit{1} & \textit{11}~~ & ~~7 &
69~~ & ~~0 & 40~~\\\hline
~$\Sigma$~ & \multicolumn{2}{c|}{} & ~~2 & 75~~ & \multicolumn{2}{c|}{} &
~~3 & 45~~\\\hline
\end{tabular}
\caption{Energies and B(E1) values for the three most dominant peaks
in the PDR area around the neutron threshold for the nucleus
$^{208}$Pb for continuum (CRPA) and discrete (DRPA) calculations.}
\label{tab2}%
\end{table}

Closer investigation of pygmy resonances have shown that this mode is
in the neighborhood of the neutron separation threshold, slightly
below for small and slightly above for large neutron excess (see for
instance Ref.~\cite{PNVR.05}). It is therefore of particular
importance to study this mode with a proper treatment of the
continuum, since in most of the previous investigations this has not
been possible. We show in Fig.~\ref{fig5} the details
of the PDR in the nucleus $^{208}$Pb. Above the theoretical neutron
separation threshold which is found at $E_{\mathrm{th}}=7.58$ MeV
(black arrow) we have a continuous red curve showing the E1 strength
distribution calculated with CRPA (units at the l.h.s) and also few
full blue vertical lines that correspond to the discrete poles of the
DRPA equations (\ref{E49}) (units at the r.h.s.) and with length
equal to the corresponding B(E1) values.

\begin{figure}[!t]
\centering
\includegraphics[width=250pt]{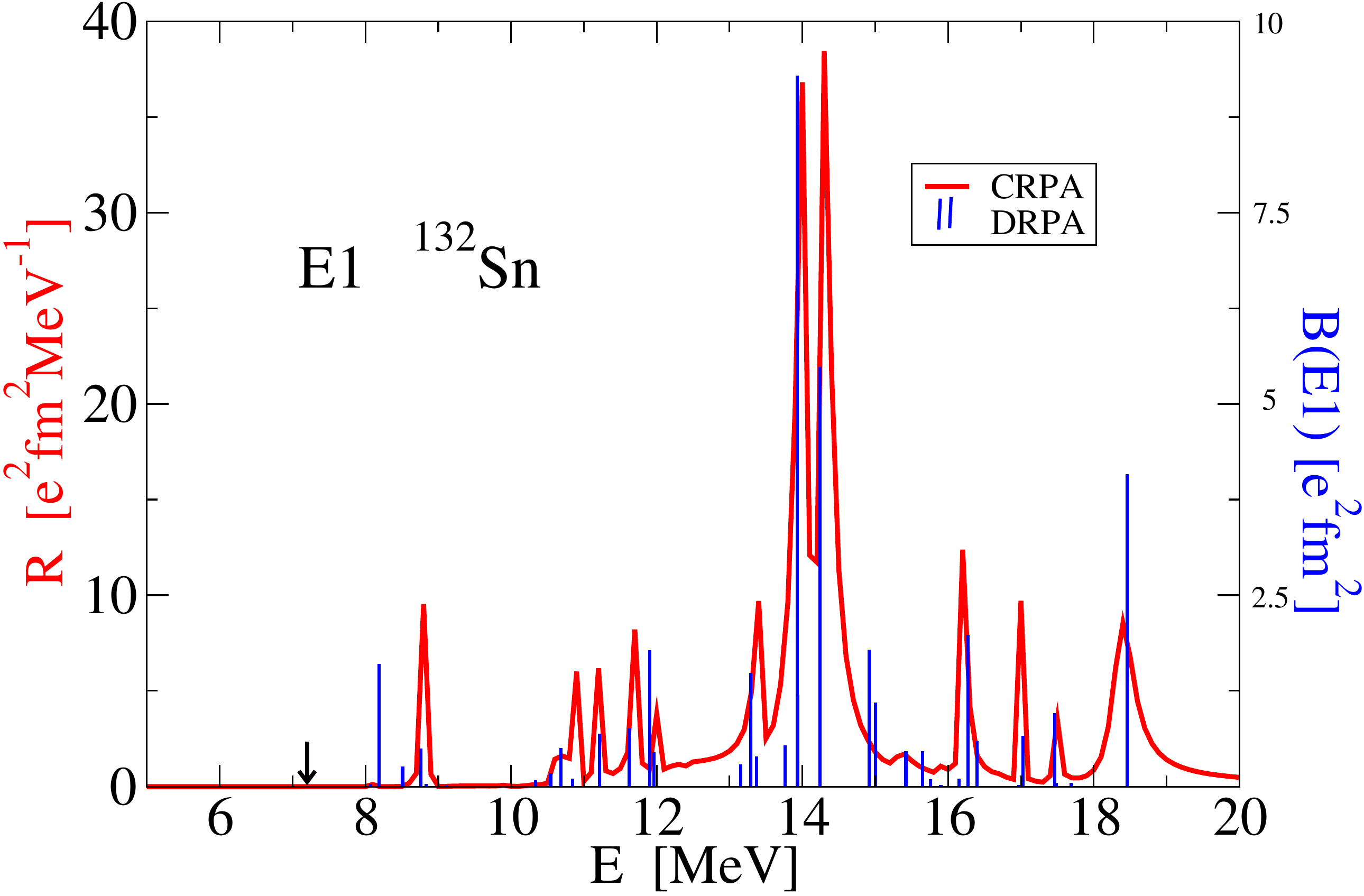}\caption{(Color online) The isovector
dipole strength distribution in $^{132}$Sn. Details are the same as
in panel (a) of Fig. \ref{fig4}.}%
\label{fig6}%
\end{figure}

In the same figure and below the threshold we have in both cases
discrete lines. The solid blue ones are again the eigen-solutions  of
the DRPA-equation (\ref{RPA-diagonalization}). The solutions of the
CRPA equations lead in this region also to discrete poles. We show
them by dashed red lines at the pole of the full response function.
Numerically, the only way to determine the B(E1) values of these
poles in CRPA is by using very small imaginary parts
$\Delta\rightarrow0$ in the frequency $\omega+i\frac{1}{2}\Delta$ and
then determining the B(E1) values by simple integration over a small
interval around this pole.

By doing that, we finally observe that there are differences in the
details between the continuum and the discrete RPA calculations close
to the neutron separation threshold. In Table~\ref{tab2} we show for
both calculations the three most dominant peaks in the area of the
PDR around $7.5$ MeV. In the discrete calculations (DRPA) the
strength is concentrated in one peak at $E=7.46$ MeV, whereas in the
continuum calculations (CRPA) most of the strength in this region is
distributed over two peaks, one below the neutron threshold at
$E=7.44$ MeV and a sharp resonance slightly above the threshold at
$E=7.66$ MeV. The energy weighted strength in this area is 17.09
[e$^{2} $fm$^{2}$] (i.e. 1.86 \% of the total sum rule) for CRPA and
26.95 [e$^{2} $fm$^{2}$] (i.e. 2.85 \% of the total sum rule) for
DRPA.

\begin{figure}[!t]
\centering
\includegraphics[width=250pt]{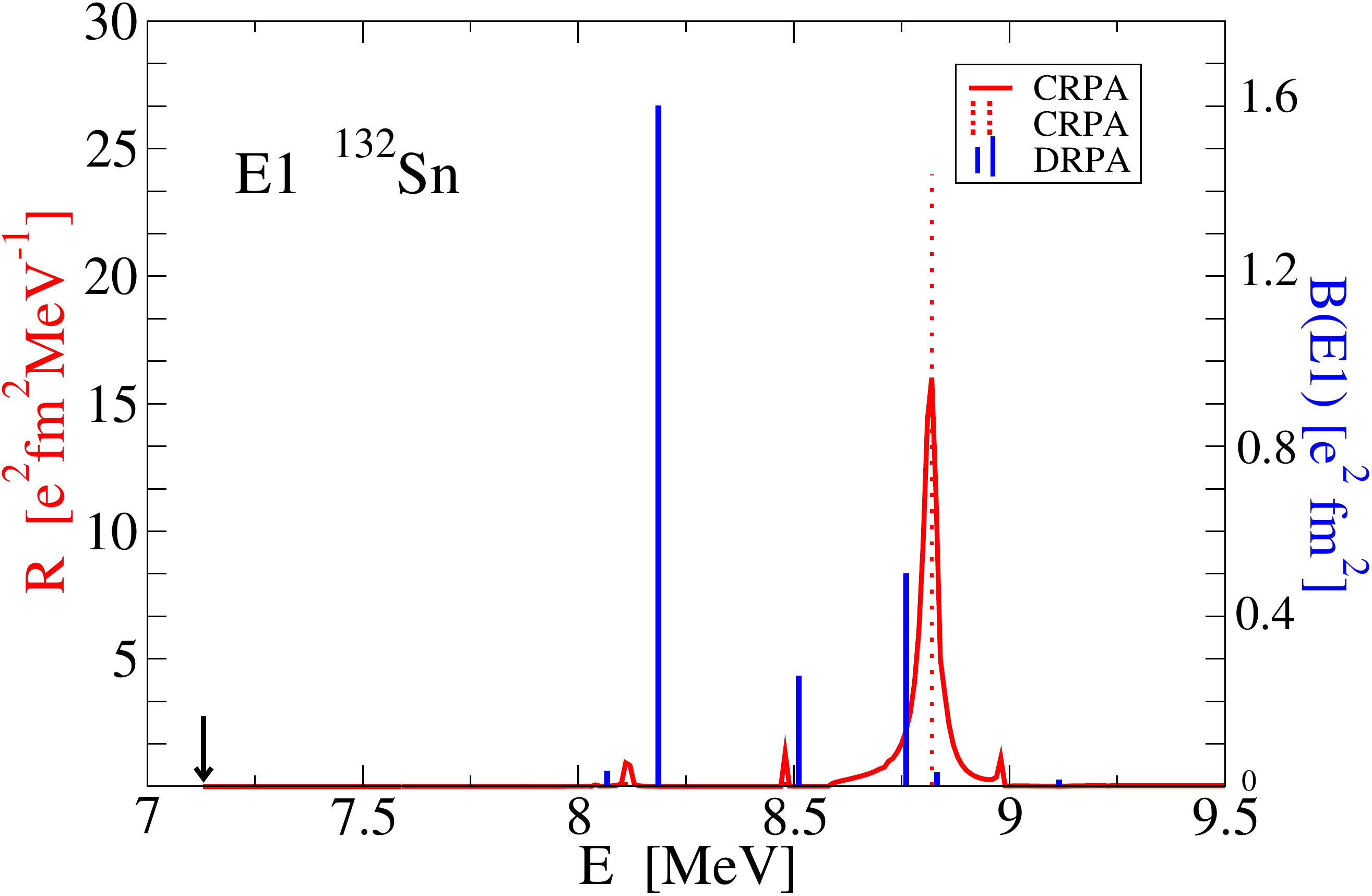}\caption{(Color online) The E1 pygmy
resonance (PDR) in the nucleus $^{132}$Sn. Details are the same as in
Fig. \ref{fig5}. The arrow indicates the theoretical neutron emission
threshold at
$E_{\mathrm{th}}=7.13$ MeV.}%
\label{fig7}%
\end{figure}

In Fig.~\ref{fig6} we show the distribution of the isovector dipole
strength in the doubly magic nucleus $^{132}$Sn. Again, results using
continuum RPA equations (red curve) are compared with the solutions
obtained from the spectral representation (blue lines). As one can
see, there is excellent agreement between the two methods, as far as
the resonance position and the overall distribution is concerned.
Moreover, the energy weighted sum rule obtained in CRPA is given by
$m_{1}(E1)=563.60$ [MeV$\cdot$fm$^{2} $], which is in very good
agreement with the DRPA calculation $m_{1}(E1)=591.02$
[MeV$\cdot$fm$^{4}$] and 22,9 \% larger than the Thomas-Reiche-Kuhn
sum rule in Eq.~(\ref{TRK})

\begin{table}[!h]
\centering
\renewcommand{\arraystretch}{1.5}%
\begin{tabular}
[c]{|c|r@{.}l|r@{.}l|r@{.}l|r@{.}l|}\hline
~~ & \multicolumn{4}{c|}{CRPA} & \multicolumn{4}{c|}{DRPA}\\\hline
~~No.~~ & \multicolumn{2}{c|}{~E~} & \multicolumn{2}{c|}{~B(E1)~} &
\multicolumn{2}{c|}{~E~} & \multicolumn{2}{c|}{~B(E1)~}\\\hline
~1~ & ~~8 & 11~ & ~~0 & 03~ & ~~8 & 067 & ~~0 & 037\\
~2~ & ~~8 & 48~ & ~~0 & 02~ & ~~8 & 186 & ~~1 & 601\\
~3~ & ~~8 & 82~ & ~~1 & 44~ & ~~8 & 511 & ~~0 & 260\\\hline
~$\Sigma$~ & \multicolumn{2}{c|}{} & ~~1 & 490 & \multicolumn{2}{c|}{} &
~~1 & 898\\\hline
\end{tabular}
\caption{Energies and B(E1) values for the three most dominant peaks
in the PDR area above the neutron threshold for the nucleus
$^{132}$Sn for continuum (CRPA) and discrete (DRPA) calculations. The
units are MeV for the energies and [e$^{2}$fm$^{2}$] for the B(E1)
values. More details are given in the text}
\label{tab3}%
\end{table}

In addition, we find that the escape width in this nucleus is
considerably smaller in the E1 channel as compared to the E0 channel
in Fig.~\ref{fig2}. This has the following explanation: The selection
rules for $ph$-excitations with E0 character is $\Delta j=0$ and no
change in parity. It turns out that most of the $ph$-excitations
contributing to the strong peak in the resonance region have rather
small $\ell$ values for the particle configurations and therefore a
very low or no centrifugal barrier. This is different for the E1
resonance, where one has a change in parity and in addition changes of
$\Delta j=0,\pm 1$. In such a case, a large part of the contributing
$ph$-pairs have particles with larger $\ell$-values i.e. a strong
centrifugal barrier and hence the width becomes smaller.

\begin{figure}[!t]
\centering
\includegraphics[width=350pt]{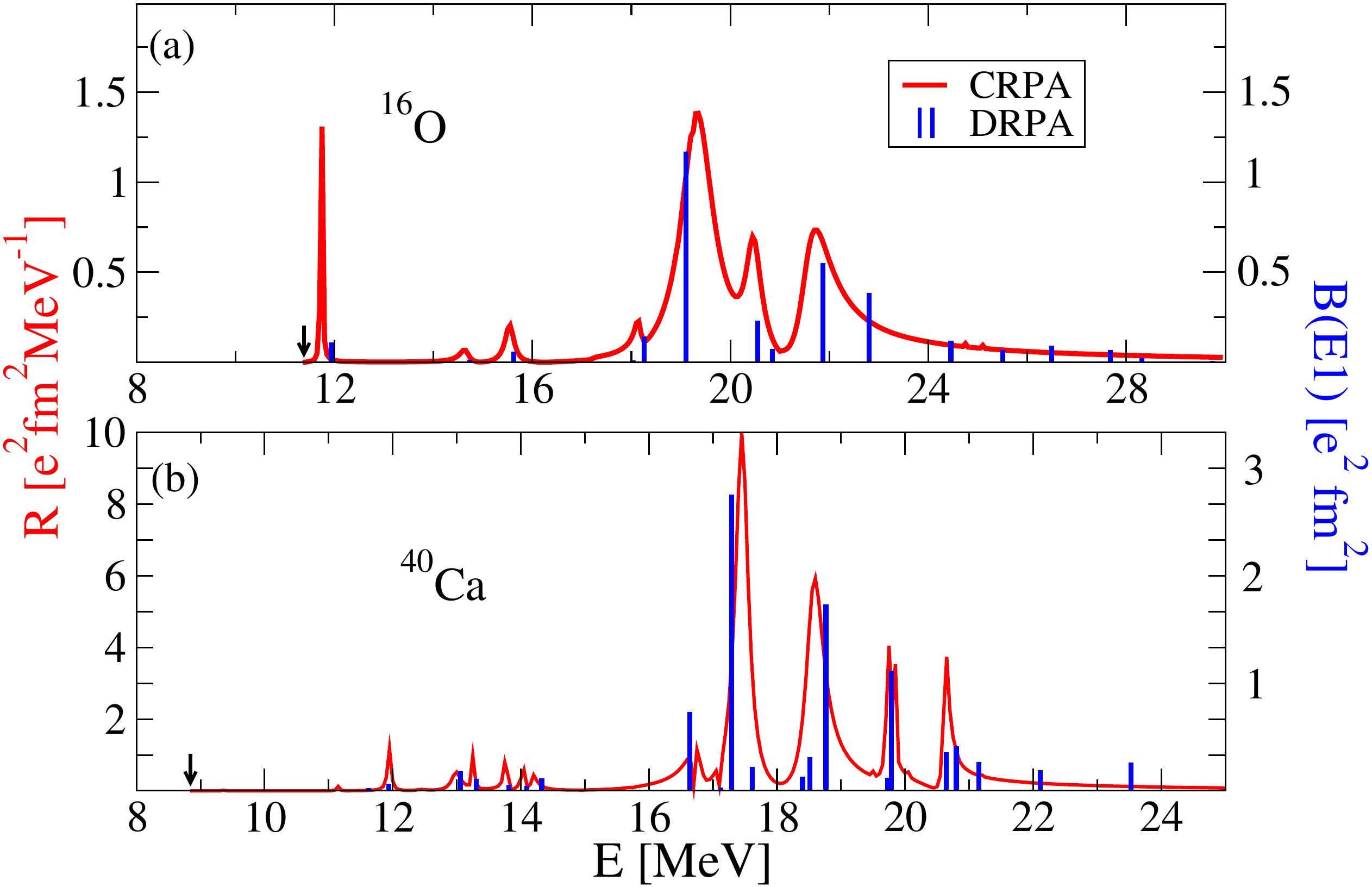}%
\caption{(Color online) The isovector dipole strength distribution in
the nuclei $^{16}$O (a) and $^{40}$Ca (b). Details are the same as in
panel (a) of Fig. \ref{fig4}. The theoretical neutron separation
energies, indicated by black arrow are $E_{\mathrm{thr}}=11.33$ MeV
for $^{16}$O and $E_{\mathrm{thr}}=8.91$ MeV for $^{40}$Ca.}%
\label{fig8}%
\end{figure}
In Fig.~\ref{fig7} we show the region of the PDR in the doubly magic
nucleus $^{132}$Sn. As already found in Ref.~\cite{PNVR.05}, the
theoretical neutron emission threshold at $E=7.13$ MeV lies much
below the area of interest. As before, we calculate the B(E1) values
of the prominent peaks, for both discrete and continuum calculations
with the total strength to be in good agreement. In Table~\ref{tab3} we show in what extent each level contributes to the
total pygmy collective state. Finally, the energy weighted strength
$m_{1}$ in this area is 13.24 [e$^{2} $fm$^{2}$] (i.e. 2.35 \% of the
total sum rule) for CRPA and 20.45 [e$^{2} $fm$^{2}$] (i.e. 3.46 \%
of the total sum rule) for DRPA.

In Fig.~\ref{fig8} we show the electric dipole strength distribution
of the lighter nuclei $^{16}$O and $^{40}$Ca. The strength obtained
in CRPA calculations (red curves) are compared with the B(E1)-values
resulting from discrete DRPA calculations (blue lines). The position
of the corresponding peaks and poles with large strength are in
rather good agreement, as explained in Table~\ref{tab4}. We find,
however, that in the continuum calculations a much larger escape
width emerges, in particular for the nucleus $^{16}$O.

\begin{table}[!t]
  \centering
\renewcommand{\arraystretch}{1.5}%
  \begin{tabular}{|r|c|c|c|}
  \hline
  & ~~CRPA~~ & ~~DRPA~~ &  ~Exp.~ \\
  \hline
  $^{16}$O &20.6279& 21.623&23.35$\pm$0.12 \cite{Va.93} \\
  $^{40}$Ca &18.367&19.32&21.76$\pm$0.11 \cite{Ve.74} \\
  $^{132}$Sn &14.503&14.78& \\
  $^{208}$Pb &13.32&13.23&13.3$\pm$0.10 \cite{RBK.93} \\
  \hline
\end{tabular}
\caption{Isovector dipole ($IVGDR$) excitation energies in [MeV] for
several spherical nuclei, calculated with both continuum and discrete
relativistic $RPA$ based on the point coupling force PC-F1.}
\label{tab4}
\end{table}

\subsection*{Isoscalar Giant Dipole Resonances}

Besides the distribution of the isovector dipole strength which is
dominated by the IVGDR in many experimental spectra, in recent years
there has also been considerable interest in measuring the isoscalar
dipole strength distribution~\cite{DGR.97,USI.04}. In a
similar way, one expects to find the ISGDR, which corresponds to a
compression wave going through the nucleus along a definite direction
and to learn from such experiments more about the nuclear
incompressibility. Relativistic calculations based on discrete
RPA~\cite{VWR.00,Pie.01} have shown that the resonance energy
of this mode is indeed closely connected to the incompressibility of
nuclear matter.

Along with this ISGDR resonance built on $3\hbar\omega$-excitations
above 20 MeV, calculations based on both relativistic~\cite{VWR.00}
and non-relativistic~\cite{CGB.00} RPA approaches have revealed a
low-lying isoscalar dipole strength in the region below and around 10
MeV. Experimental investigations with inelastic scattering of
$\alpha$-particles at small angles~\cite{CLY.01,USI.04} have also
found isoscalar dipole strength in this region. This strength has
been attributed in  Ref. \cite{VPR.02} to an exotic mode of a
toroidal motion predicted already in early theoretical investigations
on  multipole expansions of systems with currents~\cite{DC.75}
and investigated also by semiclassical methods~\cite{BMS.93}

On the theoretical point of view, there is further interest in the
isoscalar dipole mode, characterized by the quantum numbers
($J^{\pi}=1^{-},T=0$), because it contains the Goldstone mode
connected with the violation of translational symmetry in the mean
field solutions. This mode corresponds to the center of mass motion
of the entire nucleus. Because of the missing restoring force, this
mode has vanishing excitation energy. It is one of the essential
advantages of the RPA approximation, that it preserves translational
symmetry and therefore it has an eigenvalue at zero energy with the
eigenfunction given by the $ph$-matrix elements of the linear
momentum operator.

\begin{figure}[!t]
\centering
\includegraphics[width=235pt]{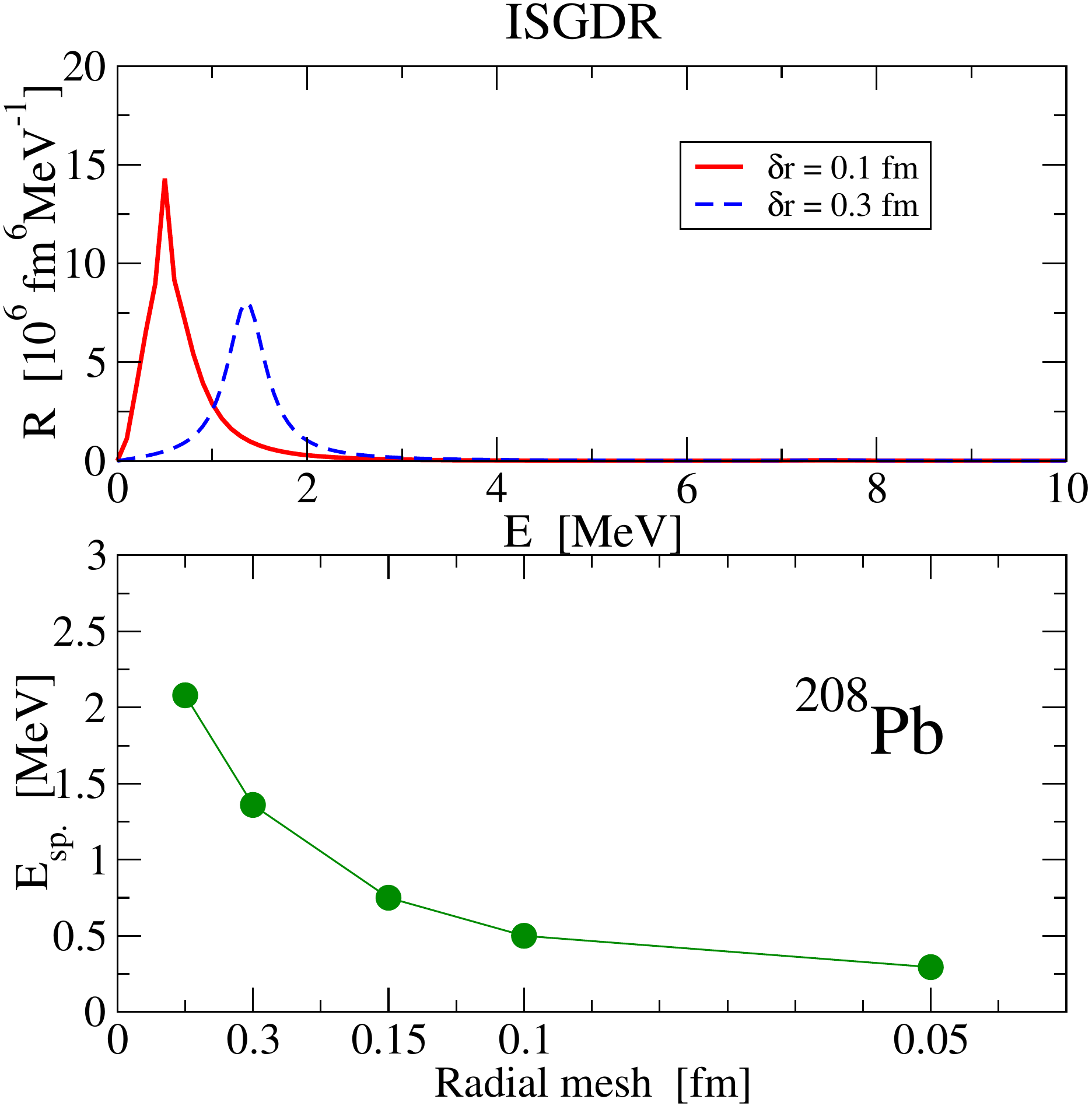}\caption{(Color online) (a)
Spurious E1 isovector strength distribution in $^{208}$Pb obtained by
CRPA calculations with two different values of the radial mesh size
$\delta r$. (b) the position of the spurious E1-state as a function
of the radial
mesh size}%
\label{fig9}%
\end{figure}
Since the ISGDR is expected to be a $3\hbar\omega$-excitation it is
usually associated with the external field derived in Ref.~\cite{SG.81}
\begin{equation}
F_{L=1}^{T=0}~=~\sum_i^A (r_i^{3} -\eta r_i ) Y_{1\mu}(\Omega_i),
\label{ISGDR-op}%
\end{equation}
where the factor $\eta=\frac{5}{3}\langle r^{2}\rangle$ is used to extract the spurious center of mass motion.

In the upper part of Fig.~\ref{fig9} we display the distribution of
the isoscalar dipole strength in $^{208}$Pb, calculated with the
operator (\ref{ISGDR-op}) for $\eta=0$, that is, we take no action for the spurious state. We therefore observe a huge peak close to zero energy, which dominates the spectrum and corresponds to the spurious translational mode.

It turns out that the position of this spurious state is an extremely
sensitive object which strongly depends on the  numerics of the
model. Of course the optimal would be to calculate the spurious state
at exactly zero energy. Therefore this excitation mode presents an
ideal benchmark for numerical efficiency of the RPA or the linear
response equations. Detailed studies have shown that the exact
separation of the spurious state requires a fully self-consistent
solution~\cite{Pie.00}; a fact which was not given in most of the
older applications with Skyrme or Gogny forces. In many cases, only
few of the different terms in the residual interaction had been taken
into account in RPA calculations.

\begin{figure}[!t]
\centering
\includegraphics[width=300pt]{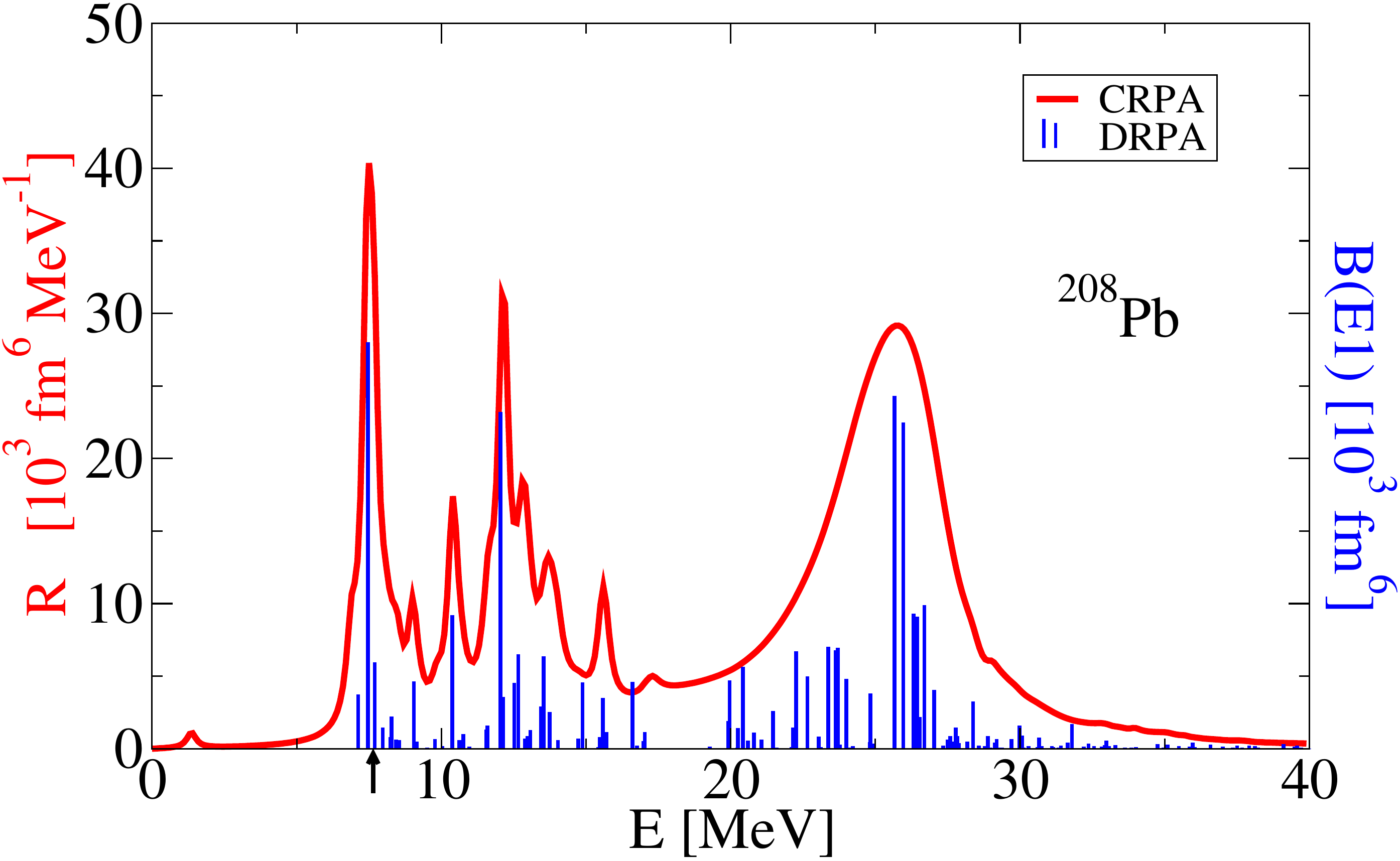}\caption{(Color online) The isoscalar
dipole strength distribution in $^{208}$Pb. Details are the same as
in the panel (a) of Fig. \ref{fig4}.}%
\label{fig10}%
\end{figure}

In addition, the configuration space must be full. Indeed, the
discussed drawback of the conventional spectral representation in a
truncated $ph$-configuration space affects the position of the
spurious state. Therefore, the convergence to zero eigenvalue of the
spurious translational mode occurs very slowly and only in extremely
large configuration space. In relativistic applications this is
translated to including also large spectrum in the Dirac
sea~\cite{DF.90,RMG.01}. As a consequence, in the spectral
representation, one has to take into account many configuration with
particles in the Dirac and holes in the Fermi sea, which complicates
the numerical applications considerably and inevitably decreases the
efficiency of the method.

Fortunately, using the continuum RPA approach, one is free from such
constraints and limitations, since the entire configuration space is
automatically included. The results in Fig.~\ref{fig9} obtained with
the operator (\ref{ISGDR-op}) for $\eta=0$ show clearly the spurious
state dominating the entire spectrum (see the scale). Its position is
not precisely at zero energy, rather it depends on the mesh size used
for the solution of the continuum response equation (the response
mesh). In panel (a) of Fig.~\ref{fig9} we present two calculations
with different mesh-sizes, where in panel (b) we show how the
spurious state moves to zero energy as we use a finer radial
interval. For the ideal case of an infinitesimal mesh, the strength
connected with the spurious state would be completely separated from
the rest of the spectrum.

In Fig.~\ref{fig10} we show results obtained with the full operator
(\ref{ISGDR-op}), i.e. with  $\eta=\frac{5}{3}\langle r^2\rangle$, in
a scale increased by three orders of magnitude. Obviously this
procedure removes the spurious state with high precision. We also did
not observe any influence of the isoscalar mode in the isovector
channel due to isospin mixing. In this context we have to remember,
that the isospin mixing introduced on the mean field level is
corrected on the RPA level to a large extend~\cite{MW.69}.

The main part of the remaining isoscalar dipole spectrum in
Fig.~\ref{fig10} is located at $E\approx25$ MeV. This "exotic" mode
is best described as a "hydrodynamical density oscillation", in which
the volume of the nucleus remains constant and the state can be
visualized as a compression wave oscillating back and forth through
the nucleus~\cite{VPR.02}.

\begin{table}[t]
\centering
\renewcommand{\arraystretch}{1.5}%
\begin{tabular}
[c]{|l|r@{.}l|r@{.}l|}\hline & \multicolumn{2}{c|}{~~Low[MeV]} &
\multicolumn{2}{c|}{~~High[MeV]}\\\hline
CRPA & ~~~~10 & 97 & ~~~~25 & 05\\
Hamamoto \textit{et al}~\cite{HSZ.98} &
\multicolumn{2}{c|}{$\sim$\,14~~~} &
23 & 4\\
Col\'o \textit{et al}~\cite{CVB.00} & 10 & 9 & 23 & 9\\
Vretenar \textit{et al}.~\cite{VWR.00} & 10 & 4 & 26 & \\
Piekarewicz~\cite{Pie.01} & \multicolumn{2}{c|}{$\sim$\,~8~~~} & 24 & 4\\
Shlomo, Sanzhur~\cite{SS.02} & \multicolumn{2}{c|}{$\sim$\,15~~~} &
\multicolumn{2}{c|}{$\sim$\,25~~~}\\\hline Uchida \textit{et
al}.~\cite{USI.04} & 12 & 7 $\pm$ 0.2 & 22 & 4 $\pm$ 0.5\\\hline
\end{tabular}
\caption{Self-consistent (relativistic and non-relativistic) RPA
calculations performed for the ISGDR in $^{208}$Pb, compared with the
most recent experimental data. The two columns refer to the centroid
energies of both the low- and high-energy sides of the ISGDR mode.}%
\label{tab5}%
\end{table}

Moreover, Fig.~\ref{fig10} shows an additional mode in the region of
$10-15$ MeV that exhausts roughly $20\%$ of the total sum rule. This
peak does not correspond to a compression mode, but as discussed in
Ref.~\cite{VPR.02} rather to a kind of toroidal motion. The toroidal
dipole mode is understood as a transverse zero-sound wave and its
experimental observation would invalidate the hydrodynamical picture
of the nuclear medium, since there is no restoring force for such
modes in an ideal fluid.

In conclusion, continuum RPA calculations manage not only to predict
the existence of the toroidal and the compression mode, but also to
achieve a reasonable agreement of the corresponding centroid energies
to other models focusing on the same problem, as well as to recent
experimental data~\cite{USI.04,DGR.97}. In Table~\ref{tab5}, these
results are presented for the case of the well studied nucleus
$^{208}$Pb.

\section{Conclusions}

\label{summary}

Starting from a point coupling Lagrangian, we have used the
non-spectral relativistic RPA approach to examine the corresponding
excitation spectra and we have compared the results with spectral
calculations based on the same Lagrangian. This non-spectral method
has several advantages. The coupling to the continuum is treated
consistently using the relativistic single particle Green's function
at the appropriate energy. In this way, complicated sums over
unoccupied states are avoided. This is particularly important for
relativistic applications since the Dirac sea is now automatically
treated properly and the unphysical transitions from holes in the
Fermi sea to particles in the Dirac sea is avoided as long as we
restrict our investigations to positive energies.

The ground state phenomena are calculated using the same Lagrangian
by a self-consistent solution of the relativistic mean field
equations in $r$-space. The residual particle-hole interaction used
in the RPA calculations is derived in a fully self-consistent way
from the second derivative of the corresponding energy density
functional. In this way no additional parameters are required and one
is able to reproduce the collective properties, namely the multipole
giant resonances for various doubly closed shell spherical nuclei
over the entire periodic table.

The calculations are carried out by using a new relativistic
continuum RPA program for point-coupling models, that includes all
the terms in the Lagrangian, in particular the two-body interactions
with zero range, the density dependent parts with all the
rearrangement terms, the derivative terms, the various
current-current terms and the Coulomb interaction. As applications
the nuclei $^{16}$O, $^{40}$Ca, $^{90}$Zr, $^{132}$Sn,
and $^{208}$Pb have been investigated proving that a hight level of
accuracy is achieved, as compared to the discrete methods. Comparing
calculations with spectral and non-spectral representations of the
response function for the same Lagrangian we find, that in general
the spectra are well reproduced within the spectral approximation, if
an appropriate phenomenological smearing parameter is used and if a
sufficiently large number of $ph$-configurations is taken into
account in the latter case. We find, however, differences in
neighborhood of the neutron threshold, where the coupling to the
continuum is not properly reproduced in the spectral method.

As compared to the discrete case the non-spectral representation has
the advantage of (i) a precise treatment of the coupling to the
continuum and a fully consistent determination of the escape width
without a phenomenological smearing parameter, (ii) a faster
evaluation of the cross section, because one needs for fixed energy
only two scattering solutions instead of the thousands of
$ph$-configurations in the discrete case and (iii) a proper treatment
of the Dirac sea without any further $ah$-configurations.

Relativistic CRPA describes very well the position of resonances in
doubly magic spherical nuclei. Provided that proper pairing
correlations are taken into account, a similar method can also be
applied in open-shell nuclei. This requires the development of the
relativistic continuum quasiparticle random phase approximation
(CQRPA). This approach accounts on equal footing for the influence of
the residual particle-hole ($ph$) as well as the particle-particle
($pp$) correlations. In analogy to non-relativistic calculations~\cite{KLL.98,HS.01,Mat.01,KSG.02} this can be achieved
on the basis of relativistic CRPA theory developed in this manuscript
either by treating the pairing correlations in the BCS approach  for
nuclei far from the drip lines where no level in the continuum is
occupied, or in the Hartree-Bogoliubov approximation valid for all
nuclei up to the drip line. Investigations in this direction are in
progress.

Of course, the present approach is based on the RPA and includes only
$1p1h$-configurations. Therefore only the escape width of the
resonances can be reproduced properly. For heavy nuclei the decay
width resulting from a coupling to more complex configurations is
very important. In fact, such couplings have been introduced
successfully in the relativistic scheme using the spectral
representation in Refs.~\cite{LRT.08}. On the non-relativistic
side, such techniques have also been used in the context of the
non-spectral representation without~\cite{KTT.97,KST.04} and
with~\cite{LT.07} pairing. So far, however, fully self-consistent
relativistic applications including complex configurations with a
proper treatment of the continuum are still missing.


Helpful discussions with G. Lalazissis, E. Litvinova, T.
Nik\v{s}i\'{c}, N. Paar, V. Tselyaev, and D. Vretenar are gratefully
acknowledged. This research has been supported the Gesellschaft f\"ur
Schwerionenforschung (GSI), Darmstadt, the Bundesministerium f\"{u}r
Bildung und Forschung, Germany under project 06 MT 246 and by the DFG
cluster of excellence \textquotedblleft Origin and Structure of the
Universe\textquotedblright\ (www.universe-cluster.de).

\begin{appendix}

\section{The effective interaction in density dependent point-coupling models}
\label{AppA}

In Eq. (\ref{energy_variation}) the effective interaction for RPA
calculations is defined as the second  derivative of the energy
functional with respect to the density matrix:
\begin{equation}
V_{\alpha\beta\alpha^{\prime}\beta^{\prime}}^{\text{ph}}=\frac{\delta
^{2}E[\hat{\rho}]}{\delta\hat{\rho}_{\alpha\beta}\delta\hat{\rho}%
_{\alpha^{\prime}\beta^{\prime}}}.
\end{equation}
In coordinate representation the indices $\alpha$,$\beta,\dots$ are
an abbreviation for the "coordinates" $1=({\bm  r_1},s_1,d_1,t_1)$,
where $s$ is the spin and $t$ the isospin coordinate, and $d=1,2$ is
the Dirac-index for large and small components. Starting from the
energy density functional (\ref{Energy}) and neglecting for the
moment the Coulomb force, we find the density dependent zero range
force
\begin{equation}
V^{\text{ph}}(1,2)~=~\sum\limits_{c}\Gamma_{c}^{(1)}~\delta({\bm r}%
_{1}-{\bm r}_{2})\upsilon _{c}({\bm r}_{1})\Gamma_{c}^{\dag(2)}%
\label{Veff}%
\end{equation}
where the \textit{vertices} $\Gamma_{c}$ are 8$\times$8 matrices
acting on the indices $s,d,t$ and reflect the different covariant
structures of the fields including spin and isospin degrees of
freedom. We express the 4$\times4$ Dirac matrices as a direct product
of spin matrices $\sigma$ and 2$\times2$ matrices $\gamma_{D}$ acting
on large and
small components%
\begin{equation}
\gamma_{0}=\left(
\begin{array}
[c]{cc}%
1 & 0\\
0 & -1
\end{array}
\right)  ,~~~1=\left(
\begin{array}
[c]{cc}%
1 & 0\\
0 & 1
\end{array}
\right)  ,~~~\gamma_{5}=\left(
\begin{array}
[c]{cc}%
0 & 1\\
1 & 0
\end{array}
\right)%
\label{gamma-2}%
\end{equation}
and the spin matrices $\sigma_{S=0}=1$ and $\sigma_{S=1}=\sigma_\mu$
with the spherical coordinates of the Pauli spin matrices. In this
way we obtain the vertices
$\Gamma_{c}=\gamma_D\times\sigma_S\times\tau_T$ as direct products of
2-dimensional Dirac-, spin- and isospin matrices (see also the second
column of Table~\ref{tab6}).

Finally, in Eq.~(\ref{Veff}) the quantities $\upsilon_{c}({\bm r}) $
describe the strengths of all the various parts of the interaction
derived in a consistent way from the Lagrangian. The ones derived
from the four-fermion terms (\ref{L_4f}) are constants. Furthermore,
due to a density dependence of the higher order terms (\ref{L_hot})
as well as the corresponding rearrangement terms, $\upsilon_{c}({\bm
r})$ depends on the static density and therefore on the coordinate
${\bm r}$. In addition, because of the derivative terms (\ref{L_der}), they also contain Laplace operators. Summarizing, we
have:
\begin{equation}%
\begin{array}
[c]{ll}%
\text{ \ \ \ }c & \text{ \ \ }\upsilon_{c}(r) =\\
\text{scalar:} & \text{ \ \
}\alpha_{S}+2\beta_{S}\rho_{S}({r})+3\gamma
_{S}\rho_{S}^{2}({r})+\delta_{S}\Delta\\
\text{time-like vector:} & \text{ \ \ }\alpha_{V}+3\gamma_{V}\rho_{V}^{2}%
({r})+\delta_{V}\Delta\text{ \ \ \ \ \ \ \ \ \ \ \ \ \ \ \ \ \ }\\
\text{space-like vector:} & \text{ \ }-\alpha_{V}-\gamma_{V}\rho_{V}^{2}%
({r})-\delta_{V}\Delta
\end{array}
\label{VphC}%
\end{equation}
In the isovector case the constants $\alpha_{S}$, $\alpha_{V}$,
$\delta_{S}$ and $\delta_{V}$ are replaced by $\alpha_{TS}$,
$\alpha_{TV}$, $\delta_{TS}$ and $\delta_{TV}$. As we see in
Table~\ref{tab1} the corresponding values $\beta_{TS}=$
$\gamma_{TS}=\gamma_{TV}$ vanish.

For spherical nuclei, the densities and currents in the Lagrangian
depend only on the radial coordinate~$r$. Therefore we expand the
$\delta$-function in Eq. (\ref{Veff}) in terms of spherical harmonics
\begin{equation}%
\delta({\bm r}_1-{\bm r}_2)~=~\frac{\delta(r_1 - r_2)}{r_1 r_2}\sum_L
Y_{L}(\Omega_{1})\cdot Y_{L}(\Omega_{2}).
\label{spher_harmon}%
\end{equation}
Combining spin ($S$) and orbital ($L$) degrees of freedom we find by
re-coupling to total angular momentum~$J$
\begin{equation}%
(\mathbf{\sigma}^{(1)}_S\cdot\mathbf{\sigma}^{(2)}_S) (Y_{L}(1)\cdot
Y_{L}(2))= \sum_J[\mathbf{\sigma} _S Y _{L}]_{J}^{(1)}
\cdot\lbrack\mathbf{\sigma} _SY _{L}]_{J}^{(2)}
\end{equation}
Inserting this expression into Eq.~(\ref{VphC}) we obtain for the
interaction a sum (or integral) of separable terms (channels)
\begin{equation}
V^{\text{ph}}(1,2)~=~\sum\limits_{c}\int\limits_{0}^{\infty}dr~Q_{c}^{(1)}%
(r)~\upsilon _{c}(r)~Q_{c}^{\dag(2)}(r)%
\label{Veff1}%
\end{equation}
Each channel is characterized by a continuous parameter $r$ and the
discrete numbers $c=(D,S,L,J,T)$. The corresponding channel operators
$Q_{c}^{(1)}(r)$ are local single particle operators
\begin{equation}
Q_{c}^{(1)}(r)~=~\frac{\delta(r-r_1)}{rr_1}\gamma^{(1)}_{D} \left[
\sigma^{(1)}_{S}Y_L (\Omega_{1})\right]  _{J}\tau^{(1)}_{T}
\label{channel-operator}%
\end{equation}
and the upper indices (1) and (2) in Eq.~(\ref{Veff1}) indicate that
these operators act on the "coordinates"
$1=(r_{1}\Omega_{1}s_{1}d_{1}t_{1})$ and
$2=(r_{2}\Omega_{2}s_{2}d_{2}t_{2})$.

The total angular momentum is a good quantum number and for fixed $J$
the sum over $c$ in Eq.~\ref{Veff1} runs only over specific numbers
$c=(D,S,L,T)$ determined by the selection rules. We concentrate in
this manuscript on states with natural parity, i.e.
$\pi=(-)^{L}=(-)^{J}$. Considering that $S=0$ for the scalar and the
time-like vector and that $S=1$ for the space-like vector we
therefore have
\begin{displaymath}
L=\left\{\begin{array}{cc} J & \text{for $S=0$} \\
                    J \pm 1 & \text{for $S=1$}
\end{array}\right.
\end{displaymath}
Finally we have eight discrete channels. Their quantum numbers are
shown in Table~\ref{tab6}.
\begin{table}[h]
\renewcommand{\arraystretch}{1.5}\centering
\begin{tabular}
[c]{|c|r@{$\,\otimes\,$}c@{$\,\otimes\,$}c|c|c|l|c|}\hline%
~c~&~$\Gamma_c=\gamma_{D}$ & $\sigma_{S}$ & $\tau_{T}$~ & ~~$D$~~ &
~~$S$~~ & ~$L$~~ & ~~$T$~~\\\hline%
~1~& $\gamma_{0}$ & 1 & 1 & $S$ & 0 & ~$J$ & 0\\
~2~& 1 & 1 & 1 & $V$ & 0 & ~$J$ & 0\\
~3~& $\gamma_{5}$ & $\sigma$ & 1 & $V$ & 1 & ~$J-1$ & 0\\
~4~& $\gamma_{5}$ & $\sigma$ & 1 & $V$ & 1 & ~$J+1$ & 0\\
~5~& $\gamma_{0}$ & 1 & $\tau_{3}$ & $S$ & 0 & ~$J$ & 1\\
~6~& 1 & 1 & $\tau_{3}$ & $V$ & 0 & ~$J$ & 1\\
~7~& $\gamma_{5}$ & $\sigma$ & $\tau_{3}$ & $V$ & 1 & ~$J-1$ & 1\\
~8~& $\gamma_{5}$ & $\sigma$ & $\tau_{3}$ & $V$ & 1 & ~$J+1$ & 1\\%
\hline
\end{tabular}
\caption{Vertices and quantum numbers of the different channels in
Eq.~(\ref{Veff})}%
\label{tab6}%
\end{table}

An essential feature of the effective interaction $(\ref{VphC})$ is
that it contains derivative terms in the form of Laplacians $\Delta$
(retardation effects are neglected). In spherical coordinates, they
contain radial derivatives as well as angular derivatives. The latter
can be expressed by the angular momentum operators acting on
spherical harmonics  $Y_{LM}$. Therefore we obtain:
\begin{equation}
\Delta=r^{2}\overleftarrow{\partial}_{r}\frac{1}{r^{2}}\overrightarrow
{\partial}_{r}+\frac{L(L+1)-2}{r^{2}}. \label{E73}%
\end{equation}
Here the radial derivatives $\overleftarrow{\partial}_{r}$ and
$\overrightarrow{\partial}_{r}$ act on the right and on the left side
in Eq.~(67),
i.e. on $\mathcal{R}_{c^{\prime}c}^{\,0}(r^{\prime}r)$ and on
$\mathcal{R}_{cc^{\prime\prime}}(r,r^{\prime \prime})$. Since the
integration is discretized $r\rightarrow r_{n}=$ $nh$ the operator
$\overrightarrow{\partial}_{r}$ is represented by a matrix in $r
$-space as for instance by the tree-point formula:
\begin{equation}
\hat{\partial}_{nn^{\prime}}=\frac{1}{2h}(\delta_{n^{\prime},n+1}%
-\delta_{n^{\prime},n-1}).
\end{equation}
This means that the term $\upsilon_{c}(r)$ in Eq. (\ref{E65}) is no
more diagonal in the coordinate $r$ and it must be replaced by a
matrix $\upsilon_{c}(r,r^{\prime})$.

The term which leads to off-diagonal terms in channel space is the
Coulomb interaction. It brakes isospin symmetry and therefore it will
be described by the general form
$\upsilon_{cc^{\prime}}(r,r^{\prime})$. In particular, we will
 have
\begin{equation}
V_{\text{C}}(1,2)=(\frac{1}{2}(1-\tau_{3} ))^{(1)}\frac{\alpha
}{|\mathbf{r}_{1}\mathbf{-r}_{2}|}(\frac{1}{2}(1-\tau_{3} ))^{(2)}%
\end{equation}
and the $r$ dependance can be written as:
\begin{equation}
\frac{\alpha}{|\mathbf{r}_{1}\mathbf{-r}_{2}|}=\sum\limits_{L}\upsilon
_{\text{C}}(r,r^{\prime})Y_{L} (\Omega)\cdot Y_{L} (\Omega^{\prime})
\end{equation}
with
\begin{equation}
\upsilon_{\text{C}}(r,r^{\prime})=\frac{4\pi\alpha}{2L+1}\cdot\frac{r_{<}^{L}%
}{r_{>}^{L+1}},
\end{equation}
and $r_{<}$ and $r_{>}$ are the smaller and the greater of $r$ and
$r^{\prime}$. This leads to a matrix $\upsilon_{cc^{\prime}}(r,r^{\prime})$ in
Eq. (\ref{E65}) as shown in Table~\ref{tab7}.%
\begin{table}[h]
\centering\renewcommand{\arraystretch}{1.5}%
\begin{tabular}
[c]{|c|cccccc|}%
\hline%
 &~~$\beta$ & $1$ & $\mbox{\boldmath$\alpha$}$ &
$\beta\vec{\tau}$ & $\vec{\tau }$ &
$\mbox{\boldmath$\alpha$}\vec{\tau}$\\%
\hline%
$\beta$ &~~0 & 0 & 0 & 0 & 0 & 0\\%
1 &~~0 & $+\frac{1}{4}\upsilon_{C}$ & 0 & 0 & $-\frac{1}{4}\upsilon_{C}$ & 0\\%
$\mbox{\boldmath$\alpha$}$ &~~0 & 0 & - $\frac{1}{4}\upsilon_{C}$ & 0 &
0 & $-\frac{1}{4}\upsilon_{C}$\\%
$\beta\vec{\tau}$ &~~0 & 0 & 0 & 0 & 0 & 0\\%
$\vec{\tau}$ &~~0 & $-\frac{1}{4}\upsilon_{C}$ & 0 & 0 & $+\frac{1}{4}%
\upsilon_{C}$ & 0\\%
$\mbox{\boldmath$\alpha$}\vec{\tau}$ &~~0 & 0 &
$-\frac{1}{4}\upsilon_{C}$ &
0 & 0 & $-\frac{1}{4}\upsilon_{C}$\\%
\hline
\end{tabular}
\caption{The structure of the channel matrix $\upsilon_{cc^{\prime}%
}(r,r^{\prime}$) for the Coulomb interaction.}%
\label{tab7}%
\end{table}

\section{The continuum representation for the Green's function}

\label{AppB} In a non-spectral or continuum approach the relativistic single
particle Green's function $G_{\kappa} (r,r^{\prime};E)$ obeys the
equation:
\begin{equation}
\left(  E-\hat{h}_{\kappa} (r)\right)%
G_{\kappa}(r,r^{\prime};E)=\delta(r-r^{\prime}),
\end{equation}
where $\hat{h}_{\kappa} (r)$ is the radial Dirac-operator of
Eq.~(\ref{Dirac-radial}) depending on the quantum number $\kappa=(lj)$. This
Green's function can be constructed at each energy $E$ from two linearly
independent solutions
\begin{eqnarray}
|u(r)\rangle&=&{\binom{f_u(r)}{g_u(r)}},\qquad%
|w(r)\rangle={\binom{f_w(r)}{g_w(r)}}\\%
\langle u^*(r)|&=&(f_u(r)\, g_u(r)),%
\langle w^*(r)|=(f_w(r)\,g_w(r))%
\end{eqnarray}
of the Dirac equation with the same energy $E$
\begin{equation}
\left(  E-\hat{h}_{\kappa} (r)\right)  |u(r)\rangle=0,\text{
\ \ \ \ }\left(  E-\hat{h}_{\kappa}(r)\right)  |w(r)\rangle=0,
\end{equation}
but with different boundary conditions.
The functions $\ u(r)$ and $w(r)$ are normalized in such a way that the Wronskian is equal to:
\begin{equation}
W=\left\vert
\begin{array}
[c]{cc}%
f_{w}(r) & f_{u}(r)\\
g_{w}(r) & g_{u}(r)
\end{array}
\right\vert =f_{w}(r)g_{u}(r)-g_{w}(r)f_{u}(r)=1.
\end{equation}
Of course these scattering solutions depend on the energy $E$ and on
the quantum number $\kappa$, i.e. we have $|u_{\kappa}(r,E)\rangle$
and $|w_{\kappa}(r,E)\rangle$. The Dirac-equation in $r$-space is a
two-dimensional equation and therefore the corresponding single
particle Green's function is a 2$\times2$ matrix. Using the bracket
notation of Dirac for the 2-dimensional spinors and following
Ref.~\cite{Tam.92} we can express this Green's function as:
\begin{equation}
G_{\kappa} (r,r^{\prime};E)=\left\{
\begin{array}
[c]{cc}%
|w _{\kappa}(r;E)\rangle\langle u^\ast_{\kappa}(r^{\prime};E)| & \text{~~for}%
\,\,r>r^{\prime}\\
|u _{\kappa}(r;E)\rangle\langle w^\ast_{\kappa}(r^{\prime};E)| & \text{~~for}%
\,\,r<r^{\prime}%
\end{array}
\right.  \label{continuum-greens-A}%
\end{equation}
with
\begin{equation}
G _\kappa(r^{\prime},r;E)= G^\top_\kappa(r,r^{\prime};E)%
\label{EB7}
\end{equation}%
The solution $u_{\kappa}(r)$ is regular at the
origin, i.e. following
Ref.~\cite{Gre.90} we have for $E>V+S$ in the limit $r\rightarrow0$:%
\begin{equation}
u(r)\rightarrow r{\binom{j_{l} (kr)}{\frac{\kappa}{|\kappa|}\frac
{E-V-S}{k}j_{\tilde{l}}(kr)}\rightarrow\binom{\frac{r}{(2l+1)!!}(kr)^{l}%
}{\frac{\kappa}{|\kappa|}\frac{r(E-V-S)}{k(2\tilde{l}+1)!!}(kr)^{\tilde{l}}}},
\end{equation}
with $k^{2}=(E-V-S)(E-V+S+2m)>0$ and $j_{l}(z)$ is a spherical Bessel
function of the first kind. The wave function $w_{\kappa}(r)$
represents at large distances for $E>0$ an outgoing wave, i.e. we
have for $r\rightarrow\infty$
\begin{equation}
w(r)\rightarrow{\binom{rh_{l}^{(1)}(kr)}{\frac{\kappa}{|\kappa|}\frac
{ikr}{E+2m}h_{\tilde{l}}^{(1)}(kr)}\rightarrow\binom{1}{\frac{\kappa}%
{|\kappa|}\frac{ik}{E+2m}}}e^{ikr},
\end{equation}
where $h_{l}^{(1)}(z)$ is the spherical Hankel function of the first kind and
for $E<0$ an exponentially decaying state, i.e. we have for $r\rightarrow
\infty$
\begin{equation}
w(r)\rightarrow{\binom{r\sqrt{\frac{2Kr}{\pi}}K_{l+\frac{1}{2}}(Kr)}%
{\frac{-Kr}{E+2m}\sqrt{\frac{2Kr}{\pi}}K_{\tilde{l}+\frac{1}{2}}%
(Kr)}\rightarrow\binom{1}{\frac{-K}{E+2m}}}e^{-Kr},
\end{equation}
where $K^{2}=(V-S-E)(E-V+S+2m)>0$ and $j_{l}(z)$ and $K_{l+1/2}(z)$
are modified spherical Bessel functions \cite{AS.70}. For $E<0$ the
two scattering solutions are both real. This absence of any imaginary
term will eventually give no contribution to the cross section of
Eq.~(\ref{strenghtfunction}). We have to keep in mind, however, that
at energies that correspond to eigen energies of a bound state, the
solutions $u_{\kappa}(r,E)$ and $w_{\kappa}(r,E)$ coincide up to a
factor, which means that the Wronskian vanishes at this energy. This
corresponds to a pole in the  response function on the real energy
axis. By adding a small imaginary part to the energy $E\rightarrow
E+i\Delta$ we  obtain a sharp peak in the strength distribution.

\section{The free response function in $r$-space}

The reduced free response\begin{equation}
R_{cc^{\prime}}^{0}(\omega)=%
\sum\limits_{ph}%
\frac{\langle h|Q_{c }^{+}|p\rangle%
\langle p|Q_{c^{\prime}}|h\rangle}%
{\omega-\varepsilon_{p}+\varepsilon_{h}}%
-\frac{\langle p|Q_{c}^{+}|h\rangle%
\langle h|Q_{c^{\prime}}|p\rangle}%
{\omega+\varepsilon_{p}-\varepsilon_{h}}%
\label{R0QQ1}%
\end{equation}
depends on the energy E and the channel indices $c,c^{\prime}$. The
operators $Q_{c}$ given by Eq. (\ref{channel-operator}) are
characterized by the channel index $c=(r,DSLT)$.  Each single
particle matrix element of the form $\langle p|Q_{c}|h\rangle$ in Eq.
(\ref{R0QQ1}) separates into an angular, an isospin and a radial
part.
\begin{equation}
\langle p|Q_{c}|h\rangle =%
\langle p|\tau_{T}|h\rangle%
\langle\kappa_{p}||\left[\sigma_{S}Y_{L}\right]_{J}||\kappa_{h}\rangle%
\langle p|\gamma_{D}|h\rangle_{r}.%
\label{QCph}
\end{equation}
Since we consider in this paper only $ph$-RPA in the same nucleus,
the particle states have the same isospin as the hole states and thus
the isospin matrix element $\langle p|\tau_{T}|h\rangle$ is simply a
phase $\pm1$.

Considering that this channel operator has a $\delta$-function in the
radial coordinate, the radial matrix elements $\langle
p|\gamma_{c}|h\rangle_{r}=\langle p(r)|\gamma_{D}|h(r)\rangle$ then
depend on $r$. They are found as
sums over the large and small components in the radial spinors
$|h(r)\rangle$ and $|p(r)\rangle$ for fixed values of $r$.

The angular matrix elements depend on the quantum numbers $\kappa$ of
particle and hole states, and, of course, on the channel quantum
numbers $S$ and $L$. In particular, we find for $S=0$:
\begin{equation}
\left\langle lj||Y_{J}||l^{\prime}j^{\prime}\right\rangle =\frac
{1+(-)^{l+l^{\prime}+J}}{2}\frac{\hat{\jmath}\hat{\jmath}^{\prime}\hat{J}%
}{\sqrt{4\pi}}(-)^{j-\frac{1}{2}}\left(
\begin{array}
[c]{ccc}%
j & J & j^{\prime}\\
-\frac{1}{2} & 0 & \frac{1}{2}%
\end{array}
\right)%
\label{yjj}%
\end{equation}
while for $S=1$, it is
\begin{eqnarray}
\label{syjj}
\langle lj||\left[\sigma Y_{L}\right]_{J}||l^{\prime}j^{\prime}\rangle &=&\frac{1+(-)^{l+l^{\prime}+L}}{2}
\frac{\hat{\jmath}\hat{\jmath}^{\prime}\hat{L}\hat{J}}{\sqrt{4\pi}}\left[(-)^{j^{\prime}+\frac{1}{2}}
\left(\begin{array}[c]{ccc} 1 & L & J\\0 & 0 & 0\end{array}\right)
\left(\begin{array} [c]{ccc} j & J & j^{\prime}\\\frac{1}{2} & 0 & -\frac{1}{2}\end{array}\right)\right. \nonumber\\
&-&\left.\sqrt{2}(-)^{l^\prime}\left(\begin{array} [c]{ccc}1 & L & J\\-1 & 0 & 1\end{array}\right)
\left(\begin{array} [c]{ccc}j & J & j^{\prime}\\\frac{1}{2} & -1 & \frac{1}{2}\end{array} \right)\right].
\end{eqnarray}

Using for the angular and isospin part the abbreviation
\begin{equation}
Q_{ph}^{c}=\langle\kappa_{p}||\left[  \sigma_{S}Y_{L}\right]
_{J}||\kappa_{h}\rangle\langle p|\tau_{T}|h\rangle,%
\label{red-me}
\end{equation}
we obtain for the reduced response function of Eq. (\ref{R0QQ}) in
$r$-space:
\begin{equation}
\mathcal{R}_{cc^{\prime}}^{0}(r,r^{\prime};\omega)  =%
\sum\limits_{ph}\left\{%
Q_{ph}^{\ast c}Q_{ph}^{c^{\prime}}%
\frac{\langle h|\gamma_{c}^{+}|p\rangle_{r}%
\langle p|\gamma_{c^{\prime}}|h\rangle_{r^{\prime}}}%
{\omega-\varepsilon_{p}+\varepsilon_{h}}%
- Q_{hp}^{\ast c}Q_{hp}^{c^{\prime}}%
\frac{\langle h|\gamma_{c^{\prime}}|p\rangle_{r^{\prime}}%
\langle p|\gamma_{c}^{+}|h\rangle_{r}}%
{\omega+\varepsilon_{p}-\varepsilon_{h}}%
\right\}%
\end{equation}%
As in Eq.~(\ref{QGQ}) we extend the sum over $p$ over the full space
and use completeness in the radial wave functions:
\begin{eqnarray}
\mathcal{R}_{cc^{\prime}}^{\,0}(r,r^{\prime};\omega)&=&{\sum\limits_{h\kappa}}\left\{Q_{\kappa h}^{\ast c}Q_{\kappa h}^{c^{\prime}}\,\langle h(r)|\gamma_{c}^{+}G_{\kappa}(r,r^{\prime};\omega+\varepsilon_{h})
\gamma_{c^{\prime}}|h(r^{\prime})\rangle\right. \nonumber \\
&+&\left.Q_{h\kappa}^{\ast c}Q_{h\kappa}^{c^{\prime}}\langle h(r^{\prime})|\gamma_{c}G_{\kappa}(r^{\prime},r;-\omega+\varepsilon_{h})
\gamma_{c^{\prime}}^{+}|h(r)\rangle\right\}.
\end{eqnarray}
Since the angular matrix elements depend only on the quantum numbers
$\kappa$ the sum over $p$ is here replaced by a  sum over the quantum
numbers $\kappa$, which is restricted by the selection rules of the
reduced matrix elements (\ref{red-me}). Having the exact form of the
Green's function for the static radial Dirac equation
(\ref{Dirac-radial}), one can finally construct the non-spectral or
continuum reduced response function (\ref{E59}):
\begin{eqnarray}
\label{E83}
\mathcal{R}_{cc^{\prime}}^{\,0}(rr^{\prime};\omega)
&=&\sum_{h\kappa}\left\{Q_{\kappa h}^{\ast c}Q_{\kappa h}^{c^{\prime}}
\gamma_{hw}^{c}(r;\omega+\varepsilon_{h})\gamma_{uh}^{c^{\prime}}(r^{\prime};\omega+\varepsilon_{h})\right.\\
&&\qquad-\left.Q_{h\kappa}^{\ast c}Q_{h\kappa}^{c^{\prime}} \gamma_{hw}^{c}(r;\omega-\varepsilon_{h})
\gamma_{uh}^{c^{\prime}}(r^{\prime};\omega-\varepsilon_{h})\right\}\mathrm{~~for~~}r>r^{\prime}\nonumber\\
&=&
\sum_{h\kappa}\left\{Q_{\kappa h}^{\ast c}Q_{\kappa h}^{c^{\prime}}
\gamma_{hu}^c(r;\omega+\varepsilon_{h})\gamma_{wh}^{c^{\prime}}(r;\omega+\varepsilon_{h})\right.\nonumber\\
&&\qquad-\left.Q_{h\kappa }^{\ast c}Q_{h\kappa}^{c^{\prime}}\gamma_{hu}^{c}(r;\omega-\varepsilon_{h})
\gamma_{wh}^{c^{\prime}}(r^{\prime};\omega-\varepsilon_{h})\right\}\mathrm{~~for~~}r<r^{\prime}\nonumber
\end{eqnarray}
where the Dirac matrix elements depend on the coordinate $r$:
\begin{eqnarray}
\gamma_{hw}^{c}(r;E) &=&\langle h|\gamma_{c}|w(E)\rangle_{r},\\
\gamma_{hu}^{c}(r;E) &=&\langle h|\gamma_{c}|u(E)\rangle_{r},\\
\gamma_{uh}^{c}(r;E) &=&\langle u^\ast(E)|\gamma_{c}|h\rangle_{r},\\
\gamma_{wh}^{c}(r;E) &=&\langle w^\ast(E)|\gamma_{c}|h\rangle_{r}.
\end{eqnarray}
Using Eq.~(\ref{EB7}) we find
\begin{equation}
\mathcal{R}_{c^{\prime}c}^{\,0}(r^{\prime},r;\omega)=
\mathcal{R}_{cc^{\prime}}^{\,0}(r,r^{\prime};\omega)
\end{equation}
It becomes clear now that the undeniable advantage of the
non-spectral approach as compared to the spectral one, is the fact
that the sum over the unoccupied states (particle states) is replaced
by a sum over the quantum number $\kappa$, which is restricted by the
selection rules for the reduced matrix elements $Q_{\kappa h}^{c}$.
For each $\kappa$, one has to determine only the pairs of the
scattering wave functions $|u\rangle$ and $|w\rangle$ for the forward
and backward term. In particular the sum over $\kappa$ does not have
to be extended over the states in the Dirac sea as in the spectral
representation (for details see Ref. \cite{RMG.01}). Therefore, not
only the size of the configuration space is significantly reduced,
but, more notably, the particle-hole as well as the antiparticle-hole
basis is taken into account fully and without any approximation.

\end{appendix}



\begin{thebibliography}{}

\bibitem{SW.86}
B.~D. Serot and J.~D. Walecka, Adv. Nucl. Phys. {\bf 16},  1  (1986).

\bibitem{BB.77}
J. Boguta and A.~R. Bodmer, Nucl. Phys. {\bf A292},  413  (1977).

\bibitem{NL3}
G.~A. Lalazissis, J. K{\"o}nig, and P. Ring, Phys. Rev. {\bf C55},  540 (1997).

\bibitem{FLW.95}
C. Fuchs, H. Lenske, and H.~H. Wolter, Phys. Rev. {\bf C52},  3043 (1995).

\bibitem{TW.99}
S. Typel and H.~H. Wolter, Nucl. Phys. {\bf A656},  331  (1999).

\bibitem{DD-ME2}
G.~A. Lalazissis, T. Nik{\v{s}}i{\'{c}}, D. Vretenar, and P. Ring, Phys. Rev.   {\bf C71},  024312  (2005).

\bibitem{VBR.95}
D. Vretenar, H. Berghammer, and P. Ring, Nucl. Phys. {\bf A581},  679 (1995).

\bibitem{RMG.01}
P. Ring, Z.-Y. Ma, N. Van~Giai, D. Vretenar, A. Wandelt, and L.-G. Cao, Nucl.   Phys. {\bf A694},  249  (2001).

\bibitem{GRT.90}
Y.~K. Gambhir, P. Ring, and A. Thimet, Ann. Phys. (N.Y.) {\bf 198}, 132   (1990).

\bibitem{ZMR.03}
S.-G. Zhou, J. Meng, and P. Ring, Phys. Rev. {\bf C68},  034323 (2003).

\bibitem{Fur.85}
R.~J. Furnstahl, Phys. Lett. {\bf B152},  313  (1985).

\bibitem{MW.69}
E.R.Marshalek and J. Weneser, Ann. Phys. (N.Y.) {\bf 53}, 569 (1969).

\bibitem{HG.89}
M. L'Huillier and N. Van~Giai, Phys. Rev. {\bf C39},  2022  (1989).

\bibitem{SRM.89}
J.~R. Shepard, E. Rost, and J.~A. McNeil, Phys. Rev. {\bf C40},  2320 (1989).

\bibitem{DF.90}
J.~F. Dawson and R.~J. Furnstahl, Phys. Rev. {\bf C42},  2009 (1990).

\bibitem{VRL.99}
D. Vretenar, P. Ring, G.~A. Lalazissis, and N. Paar, Nucl. Phys. {\bf A649},   29c  (1999).

\bibitem{VWR.00}
D. Vretenar, A. Wandelt, and P. Ring, Phys. Lett. {\bf B487},  334 (2000).

\bibitem{MGW.01}
Z. Y. Ma, N. Van~Giai, A. Wandelt, D. Vretenar, and P. Ring, Nucl. Phys. {\bf   A686},  173  (2001).

\bibitem{VPR.02}
D. Vretenar, N. Paar, P. Ring, and T. Nik{\v{s}}i{\'{c}}, Phys. Rev. {\bf C65}, 021301(R)  (2002).

\bibitem{PNVR.05}
N. Paar, T. Nik\v{s}i\'{c}, D. Vretenar, and P. Ring, Phys. Lett. {\bf B606},   288  (2005).

\bibitem{NVR.05}
T. Nik{\v{s}}i{\'{c}}, D. Vretenar, and P. Ring, Phys. Rev. {\bf C72},  014312   (2005).

\bibitem{Pie.00}
J. Piekarewicz, Phys. Rev. {\bf C62},  051304(R)  (2000).

\bibitem{Pie.01}
J. Piekarewicz, Phys. Rev. {\bf C64},  024307  (2001).

\bibitem{LRT.08}
E. Litvinova, P. Ring, and V.~I. Tselyaev, Phys. Rev. {\bf C78}, 014312 2008).

\bibitem{NVR.06}
T. Nik\v{s}i\'{c}, D. Vretenar, and P. Ring, Phys. Rev. {\bf C73}, 034308 (2006).

\bibitem{MM.89}
P. Manakos and T. Mannel, Z. Phys. {\bf A334},  481  (1989).

\bibitem{BMM.02}
T. B{\"u}rvenich, D.~G. Madland, J.~A. Maruhn, and P.-G. Reinhard, Phys. Rev. {\bf C65},  044308  (2002).

\bibitem{NVLR.08b}
T. Nik\v{s}i\'{c}, D. Vretenar, and P. Ring, Phys. Rev. {\bf C78}, 034318 (2008).

\bibitem{Wal.74}
J.~D. Walecka, Ann. Phys. (N.Y.) {\bf 83},  491  (1974).

\bibitem{NJL.61a}
Y. Nambu and G. Jona-Lasinio, Phys. Rev. {\bf 122},  345  (1961).

\bibitem{SB.75}
S. Shlomo and G.~F. Bertsch, Nucl. Phys. {\bf A243},  507  (1975).

\bibitem{BT.75}
G.~F. Bertsch and S.~F. Tsai, Phys. Rep. {\bf 18C},  125  (1975).

\bibitem{Tam.92}
E. Tamura, Phys. Rev. {\bf B45},  3271  (1992).

\bibitem{Gre.90}
W. Greiner, {\em Relativistic Quantum Mechanics} (Springer Verlag, Berlin,   1990).


\bibitem{RSp.74}
P. Ring and J. Speth, Nucl. Phys. {\bf A235},  315  (1974).

\bibitem{CG.04}
G. Col\'o and N. Van~Giai, Nucl. Phys. {\bf A731},  15  (2004).

\bibitem{YLC.04a}
D.~H. Youngblood, et.al., Phys. Rev. {\bf C69},  034315  (2004).

\bibitem{BK.47}
G.~C. Baldwin and G.~S. Klaiber, Phys. Rev. {\bf 71},  3  (1947).

\bibitem{Speth.91}
{\em Electric and Magnetic Giant Resonances in Nuclei}, edited by J. Speth   (World Scientific, Singapore, 1991), Vol.~7.

\bibitem{MDB.71}
R. Mohan, M. Danos, and L.~C. Biedenharn, Phys. Rev. {\bf C3},  1740 (1971).

\bibitem{RHK.02}
N. Ryezayeva et.al., Phys. Rev.   Lett. {\bf 89},  272502  (2002).

\bibitem{ZBH.05}
A. Zilges, M. Babilon, T. Hartmann, D. Savran, and S. Volz, Prog. Part. Nucl.
  Phys. {\bf 55},  408  (2005).

\bibitem{VPR.01a}
D. Vretenar, N. Paar, P. Ring, and G.~A. Lalazissis, Phys. Rev. {\bf
C63},
  047301  (2001).

\bibitem{RBK.93}
J. Ritman, F.-D. et. al., Phys. Rev. Lett. {\bf 70},  533  (1993).

\bibitem{Va.93}
V.V.Varlamov, Yad.Konst.,1,52 (1993)

\bibitem{Ve.74}
A.Veyssiere et. al., Nucl. Phys. {\bf A227}, 513 (1974)]  \\

\bibitem{DGR.97}
B.~F. Davis, et.al., Phys. Rev. Lett. {\bf 79},  609  (1997).

\bibitem{USI.04}
M. Uchida, et. al., Phys. Rev. {\bf C69},   051301(R)  (2004).

\bibitem{CGB.00}
G. Col{\'{o}}, N. Van~Giai, P.~F. Bortignon, and M.~R. Quaglia, Phys. Lett.   {\bf B485},  362  (2000).

\bibitem{CLY.01}
H.~L. Clark, Y.-W. Lui, and D.~H. Youngblood, Phys. Rev. {\bf C63}, 031301(R)   (2001).

\bibitem{DC.75}
V. Dubovik and A. Cheshkov, Sov. J. Part. Nucl. {\bf 5},  318 (1975).

\bibitem{BMS.93}
S.~I. Bastrukov, S. Misicu, and V. Sushkov, Nucl. Phys. {\bf A562}, 191   (1993).

\bibitem{FW.71}
A.L.Fetter, J.D.Walecka, {\em Quantum Theory of Many-Particle Systems} (McGraw Hill, New York, 1971).

\bibitem{SG.81}
N.~Van Giai and H. Sagawa, Nucl. Phys. {\bf A371},  1  (1981).

\bibitem{HSZ.98}
I. Hamamoto, H. Sagawa, and X. Z. Zhang, Phys. Rev. {\bf C57},  R1064 (1998).

\bibitem{CVB.00}
G. G.~Col\'o, N. Van~Giai, P.~R. Bortignon, and M.~R. Quaglia, Phys. Lett. {\bf   B485},  362  (2000).

\bibitem{SS.02}
S. Shlomo and A.~I. Sanzhur, Phys. Rev. {\bf C65},  044310  (2002).

\bibitem{KLL.98}
S. Kamerdzhiev, R.~J. Liotta, E. Litvinova, and V.~I. Tselyaev, Phys. Rev. {\bf   C58},  152  (1198).

\bibitem{HS.01}
K. Hagino and H. Sagawa, Nucl. Phys. {\bf A695},  82  (2001).

\bibitem{Mat.01}
M. Matsuo, Nucl. Phys. {\bf A696},  371  (2001).

\bibitem{KSG.02}
E. Khan, N. Sandulescu, M. Grasso, and N. V. Giai, Phys. Rev. {\bf C66},   024309  (2002).

\bibitem{KTT.97}
S.~P. Kamerdzhiev, G.~Y. Tertychny, and V.~I. Tselyaev, Phys. Part. Nucl. {\bf   28},  134  (1997).

\bibitem{KST.04}
S.~P. Kamerdzhiev, J. Speth, and G.~Y. Tertychny, Phys. Rep. {\bf 393},  1   (2004).

\bibitem{LT.07}
E.~V. Litvinova and V.~I. Tselyaev, Phys. Rev. {\bf C75},  054318 (2007).

\bibitem{AS.70}
M. Abramowitz and I.~A. Stegun, {\em Handbook of Mathematical Functions} (Dover Publications, New York, 1965).
\end{thebibliography}
\end{document}